\documentclass[twocolumn]{aastex631}

\newcommand{\aag}[1]{\textcolor{cyan}{#1}}

\graphicspath{{./}{figures/}}
\usepackage{graphicx}	% Including figure files
\usepackage{amsmath}	% Advanced maths commands
\usepackage{multirow}
\usepackage{soul}

\usepackage{float}
\usepackage{comment}

\newcommand{\ptcite}[1]{{\color{brown}[cite-papers: #1]}}

\begin{document}

\title{Ignition of weak interactions and r-process outflows in \emph{super}-collapsar accretion disks}

\author[0000-0002-8685-5477]{Aman Agarwal}
\affiliation{Institute of Physics, University of Greifswald, D-17489 Greifswald, Germany}

\author[0000-0001-6374-6465]{Daniel M.~Siegel}
\affiliation{Institute of Physics, University of Greifswald, D-17489 Greifswald, Germany}
\affiliation{Department of Physics, University of Guelph, Guelph, ON, N1G 2W1, Canada}

\author[0000-0002-4670-7509]{Brian D.~Metzger}
\affiliation{Columbia Astrophysics Laboratory, Columbia University, New York, NY 10027, USA}
\affiliation{Center for Computational Astrophysics, Flatiron Institute, New York, NY 10010, USA}

\author[0000-0003-2950-7772]{Chris Nagele}
\affiliation{William H. Miller Department of Physics and Astronomy, Johns Hopkins University, Baltimore, MD 21218, USA}

%\author{August Muench}
%\affiliation{American Astronomical Society \\
%1667 K Street NW, Suite 800 \\
%Washington, DC 20006, USA}

%% Note that the \and command from previous versions of AASTeX is now
%% depreciated in this version as it is no longer necessary. AASTeX 
%% automatically takes care of all commas and "and"s between authors names.

%% AASTeX 6.31 has the new \collaboration and \nocollaboration commands to
%% provide the collaboration status of a group of authors. These commands 
%% can be used either before or after the list of corresponding authors. The
%% argument for \collaboration is the collaboration identifier. Authors are
%% encouraged to surround collaboration identifiers with ()s. The 
%% \nocollaboration command takes no argument and exists to indicate that
%% the nearby authors are not part of surrounding collaborations.

%% Mark off the abstract in the ``abstract'' environment. 
\begin{abstract}
The collapse of rotating massive ($\sim\!10 M_\odot$) stars resulting in hyperaccreting black holes (BHs; ``collapsars'') is a leading model for the central engines of long-duration gamma-ray bursts (GRBs) and a promising source of rapid neutron capture (``r-process'') elements. R-process nucleosynthesis in disk outflows requires the accretion flow to self-neutronize. This occurs because of Pauli-blocking at finite electron degeneracy, associated with a critical accretion rate $\dot M > \dot{M}_{\rm ign}$. We analytically examine the assumptions underlying this ``ignition threshold'' and its possible breakdown with increasing BH mass $M_\bullet$. Employing three-dimensional general-relativistic magnetohydrodynamic simulations with weak interactions, we explore the physical conditions of collapsar accretion disks with $M_\bullet\sim 80-3000 M_\odot$ over more than a viscous timescale as they transition through the threshold. There is remarkable agreement between our simulations and the analytic result $\dot{M}_{\rm ign}\propto \alpha^{5/3}M_\bullet^{4/3}$ for $M_\bullet \sim 3-3000 M_\odot$. Simulations and analytic analyses consistently show that the largest BHs leading to r-process nucleosynthesis at $\dot{M}_{\rm ign}$ are $\approx 3000 M_\odot$, beyond which self-neutronization ceases, since the disk temperature $T\propto M_\bullet^{-1/6}$ decreases below the neutron-proton mass difference ($\simeq\!\text{MeV}$), suppressing the conversion of protons into neutrons.
%\textcolor{red}{(is this sentence referring to the simulation results or the analytic estimate. I guess its kind of both, but as written it may lead to a bit of confusion. One suggestion would be to start the sentence with ... Both the simulations and the scaling law...)}. 
We show that stellar models of $\sim\!250-10^5M_\odot$ can give rise to BHs of $M_\bullet \sim 30-1000 M_\odot$ accreting at $\dot M\gtrsim \dot{M}_{\rm ign}$, yielding $\sim\!10-100 M_\odot$ of light and heavy r-process elements per event. These rare but prolific r-process sources in low-metallicity environments are associated with super-kilonovae and likely extremely energetic GRBs. Such signatures may be used to probe Population III stars.
\end{abstract}

%% Keywords should appear after the \end{abstract} command. 
%% The AAS Journals now uses Unified Astronomy Thesaurus concepts:
%% https://astrothesaurus.org
%% You will be asked to selected these concepts during the submission process
%% but this old "keyword" functionality is maintained in case authors want
%% to include these concepts in their preprints.
%keywords{Classical Novae (251) --- Ultraviolet astronomy(1736) --- History of astronomy(1868) --- Interdisciplinary astronomy(804)}

%% From the front matter, we move on to the body of the paper.
%% Sections are demarcated by \section and \subsection, respectively.
%% Observe the use of the LaTeX \label
%% command after the \subsection to give a symbolic KEY to the
%% subsection for cross-referencing in a \ref command.
%% You can use LaTeX's \ref and \label commands to keep track of
%% cross-references to sections, equations, tables, and figures.
%% That way, if you change the order of any elements, LaTeX will
%% automatically renumber them.
%%
%% We recommend that authors also use the natbib \citep
%% and \citet commands to identify citations.  The citations are
%% tied to the reference list via symbolic KEYs. The KEY corresponds
%% to the KEY in the \bibitem in the reference list below. 

\section{Introduction} \label{sec:intro}

Identifying the astrophysical sites that host the formation of the heaviest elements in our universe, those produced through the rapid neutron capture process ($r$-process; \citealt{cameron_origin_1957,burbidge_synthesis_1957}), is a major goal in nuclear astrophysics \citep{horowitz_r-process_2019,cowan_origin_2021,siegel_r-process_2022}.  Neutron star mergers were confirmed as $r$-process sources \citep{lattimer_tidal_1976,symbalisty_neutron_1982} following the discovery of kilonova emission \citep{li_transient_1998,metzger_electromagnetic_2010,barnes_effect_2013} from the gravitational wave source GW170817 \citep{the_ligo_scientific_collaboration_search_2017}.  Several kilonovae have also been detected following GRBs from compact object mergers (e.g., \citealt{berger_r-process_2013,tanvir_kilonova_2013,Rastinejad+25}). Recently, the giant flares of magnetars, the energy released from which can excavate and eject neutron star crust material into space \citep{Cehula+24}, were also confirmed as $r$-process sources \citep{Patel+25b}, following the discovery of nuclear gamma-ray line emission after the famous 2004 flare from the Galactic magnetar SGR 1806-20 \citep{Patel+25a}.

A minimal condition for the $r$-process to take place from an explosive energetic event is that the outflowing matter be {\it neutron-rich}, i.e.~possess an electron fraction $Y_e < 0.5$ (however, see \citealt{Meyer+02} for an exception). Because the interiors of neutron stars are manifestly neutron-rich, this baseline condition is relatively easy to satisfy in the ejecta from neutron star mergers or magnetar giant flares.  By contrast, whether considering the pristine matter left over from the Big Bang, or that subsequently processed by stellar nucleosynthesis, the vast majority of the matter making up our universe has $Y_{e} \ge 0.5.$  

Nevertheless, $Y_e$ is a malleable quantity: at sufficiently high densities and temperatures, weak interactions are rapid, changing $Y_e$ from its initial value. At temperatures $k_{\rm B} T \gtrsim$ MeV the timescale for a free proton to capture an electron and become a neutron is short $\lesssim $ seconds.  If densities are furthermore also high, electrons and positrons are degenerate; under these conditions, the inverse reaction$-$positron capture by neutrons$-$is suppressed by Pauli blocking, thereby driving $Y_e$ to low values $\ll 0.5$ (e.g., \citealt{beloborodov_nuclear_2003}).  Though rare, such extreme conditions are achieved in a handful of astrophysical events which involve super-critical accretion onto a compact object \citep{chen_neutrino-cooled_2007,metzger_conditions_2008,Siegel_collapsars_2019}.  Prime examples here include post-merger accretion disks formed in the aftermath of neutron-star mergers \citep{metzger_time-dependent_2008,siegel_three-dimensional_2017,fernandez_long-term_2019,combi_jets_2023-2}, the accretion-induced collapse of a rotating white dwarf to form a rotating neutron star with an accretion disk \citep{dessart_multidimensional_2006,Metzger+09b,Cheong+25}, and the core collapse of a massive rotating star to form a spinning black hole with an accretion disk (``collapsar''; \citealt{woosley_gamma-ray_1993,macfadyen_collapsars_1999}).

In collapsars, most of the infalling stellar envelope which feeds the black hole is composed of elements with equal numbers of protons and neutrons, corresponding to $Y_e \simeq 0.5$.  The innermost regions of collapsar disks are always hot; however, the disk material will only become dense, degenerate and hence neutron-rich, if it is able to cool (lose entropy) efficiently via thermal neutrino emission \citep{beloborodov_nuclear_2003}. Because the cooling rate depends on the gas density, neutronization of the disk to $Y_{e} \ll 0.5$ only occurs above a minimum, so-called ``ignition'' accretion rate, $\dot{M}_{\rm ign}$ \citep{metzger_conditions_2008,Siegel_collapsars_2019,de_igniting_2021}. Depending on the efficiency of angular momentum transport in the disk (i.e., the effective viscosity parameter $\alpha \sim 10^{-2}-1$), the ignition rates for solar-mass black holes can range from $\dot{M}_{\rm ign} \sim 10^{-3}-1M_{\odot}$ s$^{-1}$. For $\dot{M} > \dot{M}_{\rm ign}$, the balance between turbulent heating and neutrino cooling regulates the disk to a mildly degenerate condition, with $Y_{e}$ driven to an equilibrium value $\approx 0.1-0.2$ (e.g., \citealt{siegel_three-dimensional_2017,de_igniting_2021}). 

Given also a strong magnetic field threading the black hole, the accretion rates needed for disk neutronization are similar to those required to power the relativistic jets of cosmological GRBs via the Blandford-Znajek mechanism \citep{Gottlieb+24,Issa+24}.  Similar magnetized outflows launched from the disk itself can potentially unbind neutron-rich disk material, thus enabling an $r$-process in the wind ejecta \citep{siegel_three-dimensional_2018,fernandez_long-term_2019,Issa+24,Dean&Fernandez24}.  This implicates collapsar disk winds as potentially important $r$-process sources \citep{Siegel_collapsars_2019}, though in detail the final synthesized abundances depend on the impact of neutrino absorption reactions in raising $Y_e$ of the outflowing gas \citep{pruet_nucleosynthesis_2004,surman_r-process_2008,Miller+20,li_neutrino_2021}. 

The collapsar events responsible for most long-duration GRBs likely originate from stellar progenitors with ZAMS masses $M_{\rm ZAMS} \gtrsim 40M_{\odot}$ and helium cores at death of $\lesssim 10M_{\odot}$ \citep{heger_presupernova_2005,Gottlieb+24}, as evidenced by the ejecta properties of the supernovae observed to accompany a dozen or so long GRBs \citep{woosley_supernova_2006}.  If their accretion disk outflows synthesize $r$-process elements \citep{Siegel_collapsars_2019,Miller+20,Just+22,Issa+24}, then the inner layers of these supernova ejecta will be contaminated with these ultra-heavy elements.  The high line opacities of lanthanide/actinide elements in the expanding ejecta \citep{kasen_opacities_2013} will act to trap radiation and significantly redden the supernova emission as compared to non $r$-process enriched explosions \citep{Barnes+22}.  Several searches have recently been made for such $r$-process infrared emission following the energetic supernovae which accompany GRBs \citep{Anand+24,Rastinejad+24,blanchard_jwst_2024}, though the results remain inconclusive.  

Helium cores with masses $\gtrsim 40M_{\odot}$ above those typically associated with collapsars undergo pair-instability thermonuclear explosions prior to evolving to iron core collapse \citep{barkat_dynamics_1967,woosley_evolution_2002,umeda+02,woosley_models_2007,renzo_predictions_2020,farmer_mind_2019,woosley_pair-instability_2021}, thereby precluding black hole formation within a certain stellar mass range (see \citealt{Renzo&Smith24} for a review). Yet more massive stars with $M_{\rm ZAMS} \gtrsim 260M_{\odot}$, which can evolve at low metallicity to form helium cores of mass $\gtrsim 130M_{\odot}$, experience strong photodisintegration losses after the thermonuclear explosion begins, again resulting in collapse to black holes (\citealt{Bond+84,Fryer+01,2022MNRAS_Nagele_metal_free_models}).  Thus, very massive rotating stars may also end their lives by triggering collapsar-like inflow onto massive black holes $M_{\bullet} \sim 10^{2}-10^{4}M_{\odot}$; these events are referred to as ``super collapsars'' \citep{Komissarov&Barkov10,Meszaros&Rees10,Suwa&Ioka11,Yoon+15}, because the energies of their relativistic jets $\gtrsim 10^{54}$ ergs would exceed the bulk of the GRB population.  As with pair-instability supernovae, the discovery of super-collapsars at high redshifts offers one of the only ways to directly probe the first, so-called Population III, generation of stars.

As in ordinary collapsars from low-mass black holes, if the accretion flows in super-collapsars generate neutron-rich disk winds, these neutron-rich ejecta would generate kilonova-like emission \citep{Siegel_super-kilonovae_2022}. As a result of their much greater disk wind ejecta, including up to several solar masses of $r$-process elements, such ``super-kilonovae'' transients can last weeks to months, much longer than ordinary kilonovae. They may be discovered via follow-up observations of particularly energetic GRBs, or independent of a gamma-ray trigger with infrared surveys such as those conducted by the Roman Space Telescope \citep{Siegel_super-kilonovae_2022}.

The main issue addressed in this work are the conditions for super-collapsar accretion disks to become neutron-rich, the minimal requirement for $r$-process and super-kilonovae from such events.  The collapse of very massive stars $\gtrsim 10^{2}-10^{4}M_{\odot}$ can generate much higher accretion rates onto the newly formed black hole than in ordinary collapsars (e.g., \citealt{Fryer+01}). However, analytic estimates, summarized below, predict that the ignition accretion rate also increases with the black hole mass $M_{\bullet}$, as $\dot{M}_{\rm ign} \propto M_{\bullet}^{4/3}$ \citep{metzger_conditions_2008,de_igniting_2021,Siegel_super-kilonovae_2022}.  However, this estimate rests on several assumptions which can in principle break down if $M_{\bullet}$ increases by several orders of magnitude. In the spirit of earlier parameterized disk evolution studies \citep{Siegel_collapsars_2019,de_igniting_2021}, here we present neutrino-cooled three-dimensional GRMHD simulations which elucidate the $\dot{M}_{\rm ign}$ threshold across a range of $M_{\bullet} \approx 80-3000 M_{\odot}$, thus enabling more robust conclusions about whether super-collapsars could serve as the most prodigious $r$-process events in the cosmos.

This paper is organized as follows. Section \ref{sec:theory_ign_threshold} presents a short derivation of the scaling $\dot{M}_{\rm ign} \propto M_{\bullet}^{4/3}$ of the ignition accretion with black hole mass, and discusses its assumptions as well as their potential breakdown with increasing black-hole mass. In Sec.~\ref{sec:Motivation}, we employ the semi-analytical model for the collapse of massive rotating stars of \citet{Siegel_super-kilonovae_2022}, calibrated to the present GRMHD simulations, to show that stellar models of massive and super massive stars with high compactness can give rise to black holes in mass range $\sim 30-3000\ M_{\odot}$ accreting at $\dot{M}>\dot{M}_{\rm ign}$. Section \ref{Sec:numerical_methods} presents three-dimensional GRMHD simulations of collapsar accretion disks around black holes of mass $M_{\bullet} = 80-3000\ M_{\odot}$, which initially accrete at $\dot{M}\gtrsim \dot{M}_{\rm ign}$ and subsequently transition within a viscous timescale through the ignition threshold, thereby recording in detail the physical conditions of the accretion flow at $\dot M \approx \dot{M}_{\rm ign}$. In Sec.~\ref{sec:results}, we extract a scaling law for the ignition threshold from the GRMHD simulations, compare it to analytic predictions, and explore the apparent breakdown of self-neutronization at $\approx 3000 M_\odot$. Section \ref{sec:conclusions} summarizes our results and conclusions. 

\section{Ignition threshold}
\label{sec:theory_ign_threshold}

While optically thick to photons, super-critical accretion flows onto compact objects may cool efficiently via neutrino emission at sufficiently high accretion rates $\dot M$, when the accretion flow is dissociated into a nucleon plasma \citep{popham_hyperaccreting_1999,narayan_accretion_2001-1,di_matteo_neutrino_2002,beloborodov_nuclear_2003,kohri_neutrino-dominated_2005,chen_neutrino-cooled_2007,kawanaka_neutrino-cooled_2007}. Following \citet{chen_neutrino-cooled_2007}, \citet{metzger_conditions_2008,metzger_time-dependent_2008}, \citet{de_igniting_2021} we call the critical accretion rate $\dot{M}_{\rm ign}$, above which neutrinos efficiently cool the inner accretion flow around the innermost stable circular orbit (ISCO) of a black hole, the ``ignition threshold''. Efficient cooling is defined as weak interactions, predominantly electron and positron capture ($e^- + p \rightarrow n + \nu_e$; $e^+ + n \rightarrow p + \bar{\nu}_e$), balancing viscous heating. Associated with this ignition threshold is the onset of electron degeneracy at $\dot{M}\gtrsim \dot{M}_{\rm ign}$, which leads to self-neutronization of the accretion flow \citep{Siegel_collapsars_2019}. The accreting plasma deleptonizes and turns into a neutron-rich environment as parametrized by the electron (or proton) fraction $Y_e= n_{\rm p}/n_{\rm b}$, with $n_{\rm p}$ and $n_{\rm b}$ denoting the proton and total baryon number densities, respectively.

\citet{de_igniting_2021} and \citet{Siegel_super-kilonovae_2022} derived that the ignition accretion rate scales with the black-hole mass $M_{\bullet}$ and the Shakura-Sunyaev viscosity parameter $\alpha$ \citep{shakurasunayev1973black} of the accretion flow as
\begin{equation}
    \dot{M}_{\rm ign}\propto  \alpha^{5/3} M_{\bullet}^{4/3} .
    \label{ignition_threshold_sim_scaling}
\end{equation}
We revisit this derivation here in short, focus on the assumptions made, and explore the conditions under which deviations from this scaling are expected. 

\subsection{Ignition accretion rate}

We assume a Keplerian disk in Newtonian gravity with angular velocity $\Omega \simeq \Omega_{\rm K} = (GM_{\bullet}/r^{3})^{1/2}$, midplane sound speed $c_{\rm s} = \sqrt{p/\rho}=H\Omega$, with $p$ and $\rho$ denoting the midplane pressure and density, respectively, and aspect ratio $H/r$ around a stellar-mass black hole $M_\bullet\sim 3 M_\odot$. For a derivation in the general-relativistic context, see \citet{de_igniting_2021}. We denote the gravitational radius by $r_{\rm g}=G M_\bullet/c^2$. 

Since $H/r \sim \mathcal{O}(1)$ when advective cooling dominates radiative cooling, one has $H/r \sim \mathrm{constant}$ at the transition from a neutrino-cooled disk to an advective disk. At the ignition threshold, specific viscous heating,
\begin{eqnarray}
    \dot{q}_{\rm visc} \approx \frac{9}{4} \nu \Omega^{2} \approx \frac{9}{4}\alpha r^{2}\Omega^{3}\left(\frac{H}{r}\right)^{2}, \label{eq:viscous_heating}
\end{eqnarray}
where $\nu = \alpha c_{\rm s}H \simeq \alpha r^{2}\Omega(H/r)^{2}$ is the kinematic viscosity,
is balanced by specific neutrino cooling $\dot{q}_{\nu} \propto T^{6}$ in the optically-thin limit (due to the capture of relativistic electrons and positrons on free nuclei; e.g., \citealt{qian_nucleosynthesis_1996}). This regulates the disk midplane temperature at the ignition threshold to
\begin{eqnarray}
T^{6} \propto \alpha r^{2}\Omega^{3}. \label{eq:midplane_temperature_Mign_1}
\end{eqnarray}

%At the transition between efficient neutrino cooling and adiabatic evolution, the midplane temperature $T$ is determined by balancing the specific rate of neutrino cooling $\dot{q}_{\nu} \propto T^{6}$ in the optically-thin limit (due to the capture of relativistic electrons and positrons on free nuclei; e.g., \cite{qian_nucleosynthesis_1996}), with the rate of viscous heating($\dot{q}_{\rm visc}$),
%\begin{eqnarray}
%\dot{q}_{\rm visc} \approx \frac{9}{4} \nu \Omega^{2} \approx \frac{9}{4}\alpha r^{2}\Omega^{3}\left(\frac{H}{r}\right)^{2},
%\end{eqnarray}
%where $\nu = \alpha c_{\rm s}H \simeq \alpha r^{2}\Omega(H/r)^{2}$ is the kinematic viscosity, $H$ is the disk scale-height and $r$ is the radial coordinate.  We have assumed a Keplerian disk with angular velocity $\Omega \simeq \Omega_{\rm K} = (GM_{\bullet}/r^{3})^{1/2}$, midplane sound speed $c_{\rm s} = H\Omega$ and aspect ratio $H/r$.  Since $H/r \sim \mathcal{O}(1)$ once advective cooling competes with radiative cooling, we have $H/r \sim \mathrm{constant}$ at the transition to an advective disk from an efficiently neutrino-cooled disk.  This gives,
%\begin{eqnarray}
%T^{6} \propto \alpha r^{2}\Omega^{3}. \label{eq:1}
%\end{eqnarray}

Furthermore, we assume that the midplane pressure follows $p\propto T^4$, i.e. that relativistic electron-positron pressure and radiation pressure dominate over baryon pressure near the accretion threshold. Vertical hydrostatic equilibrium then implies that 
\begin{equation}
    \text{const.} \sim \left(\frac{H}{r}\right)^{2} \approx \frac{c_{\rm s}^{2}}{r^{2}\Omega^{2}} \approx \frac{P/\rho}{r^{2}\Omega^{2}}
    \propto \frac{\alpha T^{4}r}{\Omega \dot{M}},
\end{equation}
and, therefore, 
\begin{equation} 
    T^{4} \propto \frac{\dot{M}\Omega}{\alpha r}.
    \label{eq:midplane_temperature_Mign_2}
\end{equation}
We have used that the local disk mass $M_{\rm d} \propto \rho r^{2}H$ accretes on the local viscous time $t_{\rm visc} \propto r^2/\nu \sim \alpha^{-1}\Omega^{-1}(H/r)^{-2}$, from which we find that
\begin{equation}
    \dot{M} \propto \frac{M_{\rm d}}{t_{\rm visc}} \propto \frac{\rho r^{2}H}{\alpha^{-1}\Omega^{-1}(H/r)^{-2}} \propto  \alpha \rho \Omega r^{3} \left(\frac{H}{r}\right)^{3} \label{eq:Mdot_scaling}
\end{equation}
or
\begin{equation}
    \rho \propto \frac{\dot M}{\alpha \Omega r^3}. \label{eq:midplane_density_Mign}
\end{equation}

Equations~\eqref{eq:midplane_temperature_Mign_1} and \eqref{eq:midplane_temperature_Mign_2} yield
\begin{eqnarray}
\dot{M} \propto \alpha^{5/3}r^{7/3}\Omega.
\end{eqnarray}
Upon scaling $r$ to the radius of the innermost stable circular orbit, $r_{\rm ISCO} \propto M_{\bullet}$, giving $r \propto M_{\bullet}$ and $\Omega \propto M_{\bullet}^{-1}$, we arrive at Eq.~\eqref{ignition_threshold_sim_scaling}. Normalizing to a rapidly spinning black hole with dimensionless spin $\chi_\bullet = 0.95$, one finds \citep{chen_neutrino-cooled_2007,de_igniting_2021}
% add hernandez-morales_neutrino-cooled_2025 when submitted to arXiv
\begin{equation}
    \dot{M}_{\rm ign}\approx 0.001 M_\odot \, \text{s}^{-1} \left(\frac{\alpha}{0.02}\right)^{5/3} \left(\frac{M_\bullet}{3M_\odot}\right)^{4/3}.
    \label{eq:ignition_threshold_scaling_normalized}
\end{equation}

\subsection{Breakdown of the fiducial threshold}

The above threshold calculation rests upon the assumption that the transition of an advective disk to a neutrino-cooled one occurs in a state, in which (1) radiation and relativistic $e^\pm$ pressure dominate gas pressure in the thick state near the threshold; (2) the disk is optically-thin to neutrinos;  (3) relativistic electron/positron capture on free nuclei dominates the neutrino cooling rate; (4) local turbulence (e.g., the magnetorotational instability (MRI) leading to magnetohydrodynamic turbulence, described by an effective $\alpha$ viscosity) dominates over instabilities due to self-gravity in setting the accretion rate. With an increasing black-hole mass above the fiducial value of $3M_{\odot}$, several of these assumptions could, in principle, break down.  

\subsubsection{Dominance of radiation and $e^\pm$-pressure}

Following the above scalings for a thick disk ($H/r \sim 1$), the ratio of radiation pressure and relativistic $e^\pm$-pressure to baryon pressure obeys (cf.~Eqs.~\eqref{eq:midplane_temperature_Mign_2} and \eqref{eq:midplane_density_Mign})
\begin{eqnarray}
\left.\frac{p_{\rm rad,e^{\pm}}}{p_{\rm b}}\right|_{\dot{M}_{\rm ign}} &\propto& \frac{T^{3}}{\rho} \propto \alpha^{-1/6} \Omega^{3/2}r^{5/3} \nonumber\\
&\propto&\alpha^{-1/6} M_{\bullet}^{1/6},
\label{eq:Pressure_Scaling}
\end{eqnarray}
where we have used $r \propto M_{\bullet}$ and $\Omega \propto M_{\bullet}^{-1}$ as before. For a constant value of $\alpha$, the assumption of dominance of radiation and $e^\pm$-pressure is expected to further strengthen with increasing black-hole mass.

\subsubsection{Optically-thin disk}
\label{subsec:theo_optically_thin}

The critical accretion rate above which the accretion flow becomes optically thick to neutrino emission at the ISCO is expected to scale as \citep{Siegel_super-kilonovae_2022}
% add hernandez-morales_neutrino-cooled_2025 when submitted to arXiv
\begin{equation}
	\dot{M}_{\nu}\propto \alpha M_\bullet, \label{eq:Mdot_opaque}
\end{equation}
if baryon pressure dominates in this regime ($p\propto \rho T$ and thus $T\propto  r^2 \Omega^2$ in lieu of Eq.~\eqref{eq:midplane_temperature_Mign_2}). Normalized to a rapidly spinning black hole with dimensionless spin $\chi_\bullet = 0.95$ \citep{chen_neutrino-cooled_2007}, one has
% add hernandez-morales_neutrino-cooled_2025 when submitted to arXiv
\begin{equation}
	\dot{M}_{\nu} \approx 0.01 M_\odot \, \text{s}^{-1} \left(\frac{\alpha}{0.02}\right) \left(\frac{M_\bullet}{3M_\odot}\right). \label{eq:Mdot_opaque_normalized}
\end{equation}
This results in
\begin{equation}
    \frac{\dot{M}_{\nu}}{\dot{M}_{\rm ign}} \approx 10 \left(\frac{\alpha}{0.02}\right)^{-2/3}\left(\frac{M_\bullet}{3 M_\odot}\right)^{-1/3}. \label{eq:Mnu_Mign}
\end{equation}
One might thus expect the validity of a neutrino-transparent disk at the ignition threshold to break down above a critical mass of
\begin{equation}
    M_{\bullet,{\rm opaque}} \approx 3000 M_\odot \left(\frac{\alpha}{0.02}\right)^{-2/3}.
\end{equation}

The assumption that baryon pressure dominates at the neutrino opaque threshold $\dot{M}_\nu$ is well justified for stellar-mass black holes ($\sim\!3 M_\odot$) based on GRMHD simulations \citep{li_neutrino_2021}. However, due to the increasing ratio of Eq.~\eqref{eq:Pressure_Scaling} with black-hole mass at any accretion rate, the transition to the optically thick regime at much larger black-hole masses may gradually become dominated by relativistic $e^\pm$-pressure and radiation pressure ($p\propto T^4$), in which case the critical accretion rate is given by \citep{Siegel_super-kilonovae_2022}
\begin{equation}
	\dot{M}_{\nu}\propto \alpha M_\bullet^{4/3}.\label{eq:Mdot_opaque_M43}
\end{equation}
Instead of Eq.~\eqref{eq:Mnu_Mign}, one then obtains
\begin{equation}
    \frac{\dot{M}_{\nu}}{\dot{M}_{\rm ign}} \propto \alpha^{-2/3} M_{\bullet}^{0}.
\end{equation}
We speculate that this is the case at BH masses above $\sim\!10-100 M_\odot$, such that an `avoided crossing' of $\dot{M}_\nu$ and $\dot{M}_{\rm ign}$ occurs and one does not expect the validity of a neutrino-transparent disk at the ignition threshold to break down with increasing BH mass. Whether this expectation is, in fact, satisfied needs to be addressed by explicit numerical simulations.
%\textcolor{red}{What is the mass-range validity of Equation 12? Currently this section reads as if Eq. 14 should be true, but then we say that it is not. Personally I would find this easier to follow if we stated that Eq. 12 is thought to be valid up to 1000 msun, or something similar, though I understand that there is not sharp cutoff. }

\subsubsection{Dominant cooling reactions}

The ratio of cooling via electron-positron annihilation $\dot{q}_{e^{\pm}}$ to electron/positron capture onto nuclei $\dot{q}_{eN}$ scales as \citep{qian_nucleosynthesis_1996}
\begin{equation}
    \left.\frac{\dot{q}_{e^{\pm}}} {\dot{q}_{eN}}\right|_{\dot{M}_{\rm ign}} \propto \frac{T^{3}}{\rho} \propto \alpha^{-1/6} M_\bullet^{1/6},
\end{equation}
where we have evaluated the ratio at the ignition threshold in the second step using Eq.~\eqref{eq:Pressure_Scaling}. Thus, the assumption of $\dot{q}_{eN}\gg \dot{q}_{e^{\pm}}$ is expected to weaken for large black-hole masses. To estimate when this ratio becomes unity, we use more precise numbers for the midplane temperature (Eq.~\eqref{eq:midplane_temperature_Mign_1}),
\begin{equation}
    T \approx 10\, {\rm MeV} \left(\frac{\alpha}{0.02}\right)^{\frac{1}{6}} \left(\frac{M_\bullet}{3M_\odot}\right)^{-\frac{1}{6}}\left(\frac{H}{r}\right)^{\frac{1}{3}}\left(\frac{r}{r_{\rm g}}\right)^{-\frac{5}{12}}, \label{eq:midplane_temperature_Mign_normalized}
\end{equation}
and the midplane density (Eq.~\eqref{eq:Mdot_scaling}),
\begin{eqnarray}
    \rho &\approx& \frac{\dot{M}_{\rm ign}}{6\pi \alpha \Omega r^3 (H/r)^{3}} \label{eq:miplane_density_Mign_normalized} \\
    &\approx& 9\times 10^{8}{\rm g\,cm^{-3}} \left(\frac{\alpha}{0.02}\right)^{\frac{2}{3}} \mskip-5mu\left(\frac{M_\bullet}{3M_\odot}\right)^{-\frac{2}{3}}\mskip-5mu \left(\frac{H}{r}\right)^{-3} \mskip-5mu\left(\frac{r}{r_{\rm g}}\right)^{-\frac{7}{10}}\mskip-10mu. \nonumber 
\end{eqnarray}
This leads to
\begin{eqnarray}
    \left.\frac{\dot{q}_{e^{\pm}}} {\dot{q}_{eN}}\right|_{\dot{M}_{\rm ign}}  &\approx& 0.063\left(\frac{T}{\text{MeV}}\right)^{3} \left(\frac{\rho}{10^8\text{g}\,\text{cm}^{-3}}\right)^{-1}  \\ 
    &\approx & 0.038 \left(\frac{\alpha}{0.02}\right)^{-\frac{1}{6}} \left(\frac{M_\bullet}{3M_\odot}\right)^{\frac{1}{6}}\left(\frac{H}{r}\right)^{4}\left(\frac{r}{r_{\rm g}}\right)^{-1}. \nonumber
\end{eqnarray}
Given the very sensitive nature of this ratio to the uncertain value of $H/r$ for a thick disk, it is challenging to predict exactly at which black-hole mass the transition beyond unity may take place. This requires detailed numerical simulations to explore. However, due to the very weak dependence on black-hole mass the onset should be gradual. 

\subsubsection{Neutrino cooling rates (relativistic pairs)}
\label{subsec:theo_validity_neutrino_cooling_rates}

The above expressions for neutrino cooling assume ultra-relativistic electrons ($kT \gg 2m_e c^{2} \simeq 1$ MeV). Similarly, the pair capture reaction
\begin{equation}
e^{-} + p \rightarrow n + \nu_e
\end{equation}
has a threshold energy corresponding to the proton-neutron mass difference $\Delta \equiv m_n c^{2}-m_p c^{2} \approx 1.2$ MeV.  Furthermore, free nucleons will recombine into $\alpha$-particles (or never become dissociated in the first place) around this same energy, shutting off the nucleon cooling reactions.

The midplane disk temperature at the ignition threshold scales as (Eq.~\eqref{eq:midplane_temperature_Mign_1} with $r\propto M_\bullet$, $\Omega \propto M_{\bullet}^{-1}$)
\begin{equation}
    T\big\vert_{\dot{M}_{\rm ign}} \propto \alpha^{1/6}M_{\bullet}^{-1/6},
\label{eq:temperature_scaling}
\end{equation}
and thus one expects the assumption of ultra-relativistic pairs to break down at large black-hole masses. Using more precise numbers (Eq.~\eqref{eq:midplane_temperature_Mign_normalized}), we find that the midplane temperature decreases below 1\,MeV for black-hole masses above a critical value
\begin{equation}
M_{\rm \bullet, T = 1MeV} \approx 7.5\times 10^5 M_{\odot} \left(\frac{\alpha}{0.02}\right) \left(\frac{H/r}{0.5}\right)^{2}\left(\frac{r}{r_{\rm g}}\right)^{-5/2}. \label{eq:M_1MeV_thick}
\end{equation}

A similar estimate can be obtained from the thin-disk state just above the ignition threshold. Once neutrino cooling is efficient, the disk becomes geometrically thin ($H/r < 1$) and baryon pressure-dominated, such that 
\begin{equation}
    \frac{H}{r} \simeq \frac{c_{\rm s}}{r \Omega} \approx 0.08\, \left(\frac{\alpha}{0.02}\right)^{1/10} \left(\frac{M_\bullet}{3M_\odot}\right)^{-1/10} \left(\frac{r}{r_{\rm g}}\right)^{7/20},
\label{eq:Hcool}
\end{equation}
where $c_{\rm s} = (kT/m_p)^{1/2}$ is the sound speed. This expression is obtained by balancing viscous heating $\dot{q}_{\rm visc} \propto \alpha r^{2}\Omega^3 (H/r)^{2} \propto \alpha \Omega T$ (Eq.~\eqref{eq:viscous_heating}, now allowing $H/r < 1$) with neutrino cooling $\dot{q}_{\nu} \propto T^{6}$, giving $T \propto \alpha^{1/5}M_{\bullet}^{1/10}r^{-3/10}$.  With the correct normalization, one finds
\begin{equation}
    T \simeq 4.2\,{\rm MeV}\,\left(\frac{\alpha}{0.02}\right)^{1/5} \left(\frac{M_\bullet}{3M_\odot}\right)^{-1/5}\left(\frac{r}{r_{\rm g}}\right)^{-3/10}.
\label{eq:midplane_temperature_thin}
\end{equation}
The miplane temperature thus drops below 1\,MeV above a critical black-hole mass of 
\begin{equation}
    M_{\rm \bullet, T = 1MeV} \approx 3900 M_{\odot} \left(\frac{\alpha}{0.02}\right)\left(\frac{r}{r_{\rm g}}\right)^{-3/2}. \label{eq:M_1MeV_thin}
\end{equation}

These estimates suggest that we could expect the ignition accretion rate to increase above the aforementioned analytic fiducial expectations for black-hole masses larger than $\sim \!10^{3}-10^{5}\, M_{\odot}$. Unfortunately, because of the very sensitive dependence of neutrino cooling on temperature, these analytic estimates are highly uncertain. Additional uncertainty enters through the sensitive dependence of $M_{\rm \bullet, T = 1MeV}$ at the ignition threshold on the uncertain value of $H/r$ of a thick disk (Eq.~\eqref{eq:M_1MeV_thick}), which makes the estimate \eqref{eq:M_1MeV_thin} perhaps more reliable. Regardless of where the transition actually occurs as a function of $M_\bullet$, it should be gradual, because the cooling rates are skewed towards the most energetic electrons and positrons in the Fermi-Dirac distribution. This is due to the sensitive energy dependence ($\sigma \propto E^{2}$) of the weak-interaction cross section with energy $E$. These considerations suggest that detailed numerical simulations are required to explore the breakdown regime of the fiducial ignition threshold.

\section{Super-collapsars}
\label{sec:Motivation}

The expected ignition threshold (Eq.~\eqref{eq:ignition_threshold_scaling_normalized}), normalized to black-hole masses of interest here as well as effective $\alpha$-viscosities extracted from the GRMHD simulations of accretion disks around such black holes (Sec.~\ref{sec:accretion_tvisc}), reads
\begin{equation}
    \dot{M}_{\rm ign}\approx  0.15 \left(\frac{\alpha_{\mathrm{eff}}}{9\times 10^{-3}}\right)^{5/3} \left(\frac{M_{\bullet}}{80 M_\odot}\right)^{4/3} M_\odot {\rm s}^{-1} .
    \label{eq:ignition_threshold_formula_normalized_80Msun}
\end{equation}
In the following, we illustrate that the collapse of rotating massive and supermassive stars can indeed give rise to collapsar-like disks with accretion regimes $\dot M > \dot{M}_{\rm ign}$ onto black holes of mass in the range $\sim\!30-3000\,M_\odot$. 

\begin{figure}
    \centering
    \includegraphics[width=1.02\linewidth]{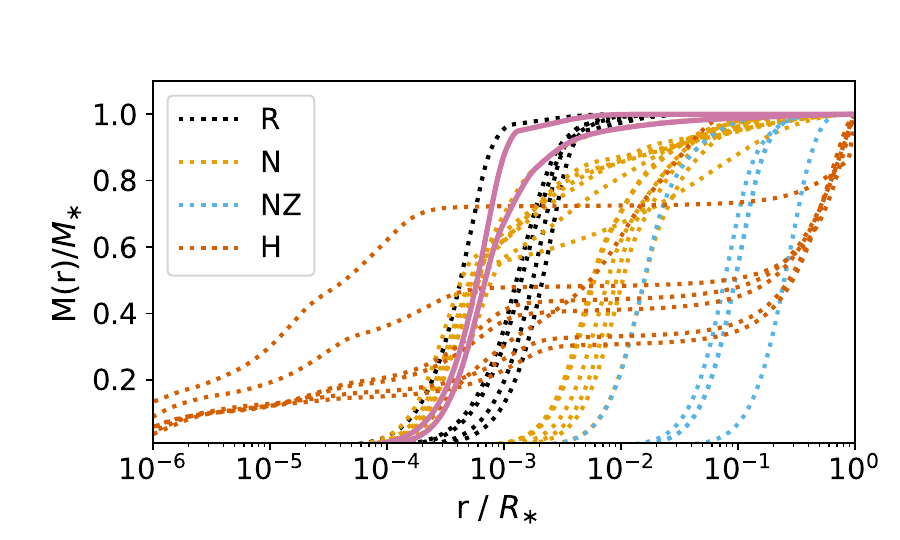}
    \caption{Enclosed fraction of stellar mass $M_\ast$ as a function of radius (in units of the stellar radius $R_{\ast}$) for various stellar models at the onset of core collapse, including the $\gtrsim\!130M_\odot$ models of \cite{renzo_predictions_2020} (R), the metal-free SMS models of \citet{2022MNRAS_Nagele_metal_free_models} (N), as well as metal-rich SMS models of \citet{nagele_evolution_2023} (NZ). For reference, progenitor models of ordinary collapsars of \citet{heger_presupernova_2000} (H) are also shown. %The models labelled 'R'($M_{\rm |collapse}\sim 10^2 M_{\odot}$) and 'N'($M_{\rm |collapse}\sim 10^4 M_{\odot}$) are very compact with almost all of the mass lying within $1\%$ of the stellar radii. The models labelled 'NZ' are similar in mass to 'N' but are metal-rich and less compact comparatively. The 'H' label models ($M_{\rm |collapse}\sim 10 M_{\odot}$) are the least compact ones with an increasing enclosed mass distribution over all radius scales. 
    The solid lines represent our fiducial models. }
    \label{fig:stellar_models}
\end{figure}

As reference scenarios in the regime $M_\bullet \approx 30-100M_\odot$, we consider the fate of massive stars (MS) with zero-age main sequence (ZAMS) mass of $M_{\rm ZAMS}\gtrsim 250 M_\odot$, which are predicted to evolve into helium cores by the time of core collapse above the pair-instability supernova (PISN) mass gap $\gtrsim 130M_\odot$ (e.g., \citealt{woosley_evolution_2002,woosley_pulsational_2017,farmer_constraints_2020,renzo_predictions_2020,woosley_pair-instability_2021}). If such stars are rapidly rotating at core collapse, they can form collapsar-like accretion disks and, as a result of mass loss through disk winds, populate the PISN mass gap with black holes in the range $\sim\!60-130\, M_\odot$ \citep{Siegel_super-kilonovae_2022}. We employ the stellar models of \citet{renzo_predictions_2020}, which were computed with the MESA stellar evolution code. These stellar models start from naked helium cores of metallicity $Z=0.001$ and are self-consistently evolved from helium core ignition to the onset of core collapse using a 22-isotope nuclear reaction network, albeit without rotation.

As reference scenarios for accreting black holes with $M_\bullet\sim\!100-1000 M_\odot$, we employ supermassive star (SMS) progenitors, which may collapse through the general-relativistic radial instability \citep{chandrasekhar_dynamical_1964}. Supermassive protostars are thought to be born rapidly rotating \citep{Kimura2023ApJ...950..184K}. Under the assumption of a constant accretion rate, deuterium, hydrogen, and eventually CNO burning set in during the accretion phase \citep{Hosokawa2013ApJ...778..178H}. Photospheric feedback is too weak to disrupt the accretion process \citep[unless the flow is variable, ][]{Sakurai2015MNRAS.452..755S}, which continues until the star collapses to a black hole or the gas reservoir is exhausted. If the latter, the SMS contracts and undergoes more or less standard massive star hydrogen and helium burning phases \citep{fuller_evolution_1986,Woods2020MNRAS.494.2236W}. These SMSs are the most relevant to the current study as they have both relatively compact cores and lower threshold masses for collapse than their accreting cousins \citep{Nagele2024PhRvD.110c1301N}. We consider both metal free \citep{2022MNRAS_Nagele_metal_free_models} and metal rich \citep{nagele_evolution_2023} progenitors where the time of collapse has been determined by post-processing stellar evolution simulations (HOSHI code \citep{Takahashi2016MNRAS.456.1320T}, with post-Newtonian gravity and $>50$ isotopes) using a normal mode analysis of adiabatic pulsations of the stars in spherical general relativity \citep{Haemmerle2021A&A...647A..83H,2022MNRAS_Nagele_metal_free_models}. This is a necessary step given the poor time resolution of the evolutionary calculations for the marginally stable SMSs. The reason that SMSs are always close to instability is that they are supported almost exclusively by radiation pressure \citep{fuller_evolution_1986}. It is thus thought that supermassive star cores should resemble $n=3$ polytropes. Since the radiation pressure is so large, the $\Omega \Gamma$ limit \citep{Maeder2000A&A...361..159M} enforces a maximum rotation rate on the SMSs \citep{Haemmerle2018ApJ...853L...3H}, and realistic rotational profiles may be even slower than this. Nevertheless, the study of rapidly rotating collapsing SMSs is well motivated given the expected rapid rotation of the progenitors in combination with the angular momentum of the accreted material, and the diversity of interesting phenomena which might originate from such a collapse. In addition to super-kilonovae \citep{Siegel_super-kilonovae_2022}, other scenarios have been considered, including ultra-long GRBs \citep{Sun2017PhRvD..96d3006S}, `torus-bounce' explosions \citep{Uchida2017PhRvD..96h3016U,Fujibayashi2024arXiv240811572F}, and gravitational wave emission \citep{reisswig_formation_2013,Shibata2016PhRvD..94b1501S}.
%Specifically, we consider the case of SMSs of $\gtrsim 10^4-10^5M_\odot$, which are expected to collapse pre-ZAMS in their protostar phase or during hydrogen burning when they are still rapidly rotating \citep{fuller_evolution_1986,haemmerle_rotation_2018,nagele_evolution_2023}. We employ both metal-free and metal-rich progenitor models (up to solar-like metallicity) of \citet{2022MNRAS_Nagele_metal_free_models} and \citet{nagele_evolution_2023}, respectively. \dms{perhaps more details here on SMS models, protostar structure, rotation, GR instability etc.}

Figure~\ref{fig:stellar_models} compares the cumulative compactness of the aforementioned sets of stellar models for MSs and SMSs at the onset of core collapse. Additionally, progenitor models of ordinary collapsars of \citet{heger_presupernova_2000} are shown for reference, which lead to black holes accreting in the $M_\bullet\approx 1-10 M_\odot$ regime. Whereas the metal-rich SMS models are among the least compact ones, many metal-free SMS models reach a compactness similarly high as the MS models. The latter models even reach higher compactness relative to the ordinary collapsar progenitor models in the stellar core. Since high core compactness is favorable for generating large accretion rates above the ignition threshold upon collapse, we focus here on typical high-compactness stellar models for purposes of further illustration. Specifically, as the fiducial MS model we employ one of the fiducial models discussed by \citet{Siegel_super-kilonovae_2022}, the model of \citet{renzo_predictions_2020} with initial helium core mass of $M_{\rm He} = 250.25 M_\odot$ and metallicity $Z=0.001$. As the fiducial SMS model, we employ the metal-free $M=3.4\times 10^4 M_\odot$ model of \citet{2022MNRAS_Nagele_metal_free_models}. Both fiducial models are indicated by solid lines in Fig.~\ref{fig:stellar_models}. 

Lacking detailed knowledge about the rotation profiles, we assume rigid rotation of spherical shells and endow the stellar models at the time of core-collapse with a parametrized specific angular momentum profile of the form (\citealt{Siegel_super-kilonovae_2022})
\begin{equation}
	j(r) = \left\{ \begin{array}{cc}
		f_{\rm K} j_{\rm K}(r)\left(\frac{r}{r_{\rm b}}\right)^{p}, & r < r_{\rm b} \\
		f_{\rm K} j_{\rm K}(r), & r_{\rm b} \le r \le R_\star
		\end{array} \right., \label{eq:j_profile}
\end{equation}
where $r_{\rm b}$, $p$, and $f_{\rm K}$ are free parameters, $j_{\rm K}(r)$ is the local Keplerian angular momentum, and $R_\star$ denotes the radius of the star. We use the semi-numerical, one-dimensional fallback model of \citet{Siegel_super-kilonovae_2022} to describe the stellar collapse, the formation of the black hole, its accretion disk as well as disk outflows, obeying mass and angular-momentum conservation.

\begin{comment}
Following \citet{Siegel_super-kilonovae_2022}, we define the following accretion regimes of the collapsar disk:
\begin{equation}
	\dot{M} = \left\{ \begin{array}{cc}
		> \dot{M}_{\nu} & \text{limited $r$-process,}\\
		& (69 \le A \le 136) \\
		\in [2\dot{M}_{\rm ign}, \dot{M}_{\nu}] & \text{main $r$-process,}\\
		& (69 \le A) \\
		\in [\dot{M}_{\rm ign}, 2\dot{M}_{\rm ign}] & \text{limited $r$-process,}\\
		& (69 \le A \le 136) \\
		< \dot{M}_{\rm ign} & \text{no $r$-process,} \\
		&  ^{56}\text{Ni production}
	\end{array}\right. . \label{eq:accretion_regimes}
\end{equation}
Here, $A$ denotes the mass number of the nuclei synthesized and $\dot{M}_{\nu}$ is the neutrino-opaque threshold (cf.~Sec.~\ref{subsec:theo_optically_thin}).
\end{comment}

\begin{figure}[t]
    \centering
    \includegraphics[width=1.0\linewidth]{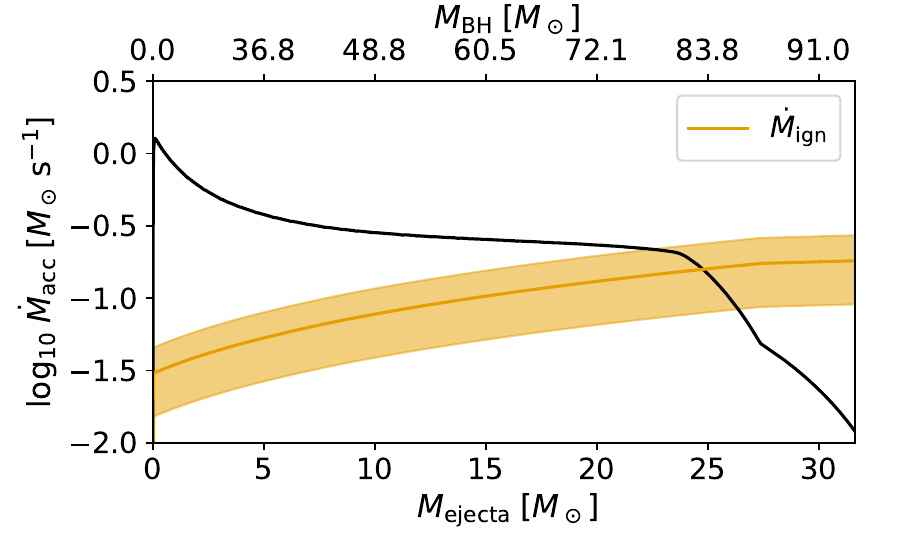}\\
    \includegraphics[width=1.0\linewidth]{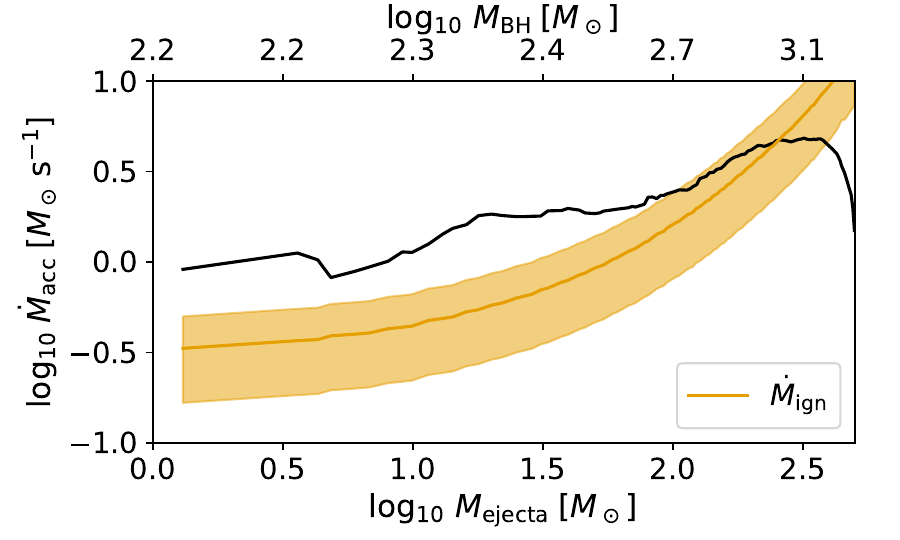}\\
    
    \caption{Top: accretion rate as a function of cumulative disk outflow mass for the $250.25 M_{\odot}$ MS model of \citet{renzo_predictions_2020} with $p=4.5$,
  $r_{\rm b}=1.5\times 10^{9}$\,cm, $f_{\rm K}=0.3$
  (see Eq.~\eqref{eq:j_profile}), and $\alpha = 0.009$. Bottom: same as top panel, but for the metal-free $3.4 \times 10^4 M_{\odot}$ SMS model of \citet{2022MNRAS_Nagele_metal_free_models}, with
  $f_{\rm K}=0.05$, $r_{\rm b}=1.9\times 10^{11}$\,cm and all other parameters identical to those above. For orientation, the instantaneous black-hole mass is also indicated at the top of the panels. The orange lines represent the ignition threshold (Eq.~\eqref{eq:ignition_threshold_formula_normalized_80Msun}), calibrated to the GRMHD simulations in this paper, with the shaded region representing the normalization uncertainty as extracted from the simulation data. The accretion disks are predicted to reside above ignition while black hole masses are in the range $\approx\!25-80 M_\odot$ and $\approx\!200-1000 M_\odot$ for the MS and SMS case, respectively. The accretion rates are time-averaged over five consecutive time steps.
  }
    \label{fig:massive_Star_fallback}
\end{figure}

Figure \ref{fig:massive_Star_fallback} shows the results of the fallback model for the aforementioned super-collapsar scenarios. The ignition threshold has been adjusted to the normalized scaling \eqref{eq:ignition_threshold_formula_normalized_80Msun}, which is calibrated to the GRMHD simulation results presented in this paper. If the outer part of the MS model rotates at 30\% of the Keplerian angular momentum, a compact accretion disk forms when the black hole has grown to $M_\bullet\approx 25 M_\odot$. The final black-hole mass is $M_\bullet\approx 80M_\odot$. The accretion rate onto the black hole exceeds $\dot{M}_{\rm ign}$ for almost the entire accretion process, resulting in a predicted synthesis of up to $\sim\!10 M_\odot$ of r-process material in disk outflows. In the case of the SMS model, the outer part of the protostar rotates at $5\%$ of the Keplerian angular momentum, which is comfortably allowed by the $\Omega\Gamma$ limit \citep{Haemmerle2018ApJ...853L...3H}. This leads to the formation of a $\approx\!200 M_\odot$ black hole, which quickly accretes at $\dot{M}>\dot{M}_{\rm ign}$ until the black hole has reached a mass of $\approx\!1000 M_\odot$ to produce up to $\sim\!20-50 \ M_\odot$ of r-process ejecta as part of the disk outflows. Depending chiefly on compactness and rotation rate, other models lead to similar results.

\begin{comment}
\begin{figure}[t]
    \centering
    \includegraphics[width=1.0\linewidth]{figures/_Mwind|alpha=0.009|p_exp =4.5| j_shell= 0.3j_kep | R_shell= 1.5e+09.pdf}\\
    \includegraphics[width=1.0\linewidth]{figures/_Mwind_alpha=0.009_p_exp =4.5_ j_shell= 0.3j_kep _ R_shell= 1.2e+12.pdf}
    \caption{Top and Bottom: the accretion rate onto the black hole as a function of ejected disk wind mass for the models presented in the top and bottom panel of Fig.~\ref{fig:vms_fallback}, respectively. While the accretion rate surpasses the ignition threshold into the blue shaded areas (cf.~Eq.~\eqref{eq:accretion_regimes}) r-process nucleosynthesis may proceed in the ejecta and produce up to $\sim\!10 M_\odot$ of r-process material in both cases.}
    \label{fig:vms_accretion_rate}
\end{figure}

\end{comment}

%Hence we investigate with state-of-the-art GRMHD simulations
\section{Numerical Methods}\label{Sec:numerical_methods}

\subsection{GRMHD simulation setup}

The simulations of accretion flows around massive black holes are performed in ideal GRMHD and full 3D, without the use of spatial symmetries. The spacetime is fixed for computational efficiency---we employ a 3+1 slicing of spacetime resulting from the horizon-penetrating Kerr-Schild coordinates. We employ the evolution code as described in more detail in \citet{siegel_three-dimensional_2018}; it represents an evolved version of \texttt{GRHydro} \citep{mosta_grhydro_2014} and makes use of the \texttt{Einstein Toolkit}\footnote{http://einsteintoolkit.org} \citep{goodale_cactus_2003,schnetter_evolutions_2004,thornburg_fast_2004,yosef_zlochower_2022_6588641,loffler_einstein_2012} as well as of \texttt{GRMHD\_con2prim} \citep{siegel_grmhd_con2prim_2018} with improved methods for the conversion of conservative to primitive variables of the GRMHD conservation equations \citep{siegel_recovery_2018}. We use the Helmholtz equation of state (EOS) \citep{ timmes_accuracy_1999, timmes_accuracy_2000} as implemented into \texttt{GRHydro} in \cite{siegel_three-dimensional_2018} to describe the thermodynamic state of the GRMHD plasma. This EOS is formulated in terms of the Helmholtz free energy, considering a multi-component ideal fluid with Coulomb corrections, electron-positron interactions, and photons in thermodynamic equilibrium. Here, the disk plasma is composed of free neutrons, free protons, and $\alpha$-particles in nuclear statistical equilibrium. Weak interactions and neutrino cooling are treated as in \cite{siegel_three-dimensional_2018}, with an energy-averaged (gray) leakage scheme that closely follows the implementation of \cite{radice_dynamical_2016}, which, in turn, is based on \citet{galeazzi_implementation_2013}, \citet{ruffert_coalescing_1996}, and \citet{Bruenn_stellar_core_collapse}. The procedure to calculate neutrino optical depths for neutrino cooling is adapted from that of \citet{Neilsen_2014_magnetized_neutron_stars}, which is well-suited for the non-spherical geometry of accretion disks. Since we are solely interested in accretion disks close to the ignition threshold of weak interactions in this paper, we neglect neutrino absorption and neutrino transport effects for computational efficiency, though we have the capability to include approximate neutrino transport via a one-moment (M0) or two-moment (M1) approximation of the general-relativistic Boltzmann equation \citep{combi_grmhd_2023,li_neutrino_2021}. At the ignition threshold, such accretion disks are optically thin to neutrinos, and transport effects are expected to be negligible. Specifically, neutrino absorption may have a minor effect on the precise composition of the outflows (distribution of proton fraction $Y_e$; \citealt{siegel_three-dimensional_2018}) and thus on precise predictions for nucleosynthesis, which, however, is not the focus of this paper.

\begin{figure*}[t]
\centering
    \includegraphics[width=1.0\textwidth]{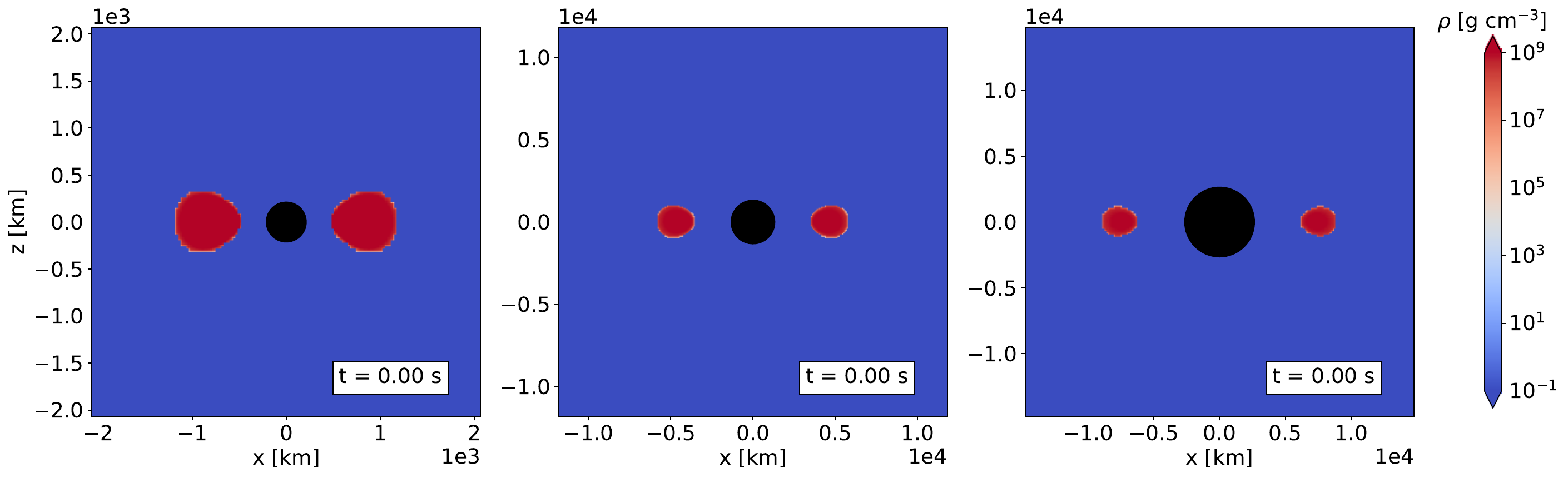}
\caption{ Sections through the initial black-hole--torus setups for simulations M80 (left), M500 (center), and M1000 (right), showing the rest mass density. The central black hole is represented by a black disk extending out to the black hole horizon (see Tab.~\ref{tab:sims_info} for parameters). The M3000 configuration (not shown here) is very similar to the M1000 setup.}
\label{fig:initial_Setup}
\end{figure*}

\begin{table*}[t]
    \centering
    \caption{Characteristics of the initial data for the fiducial accretion-disk configurations considered in this paper. From left to right: mass $M_\bullet$ of the central black hole, radius $r_{\rm H}$ of the corresponding horizon, initial inner and outer torus radius ($r_{\rm in}$ and $r_{\rm out}$), mass of the initial torus ($M_{\mathrm{disk},0}$), maximum initial density $\rho_{\mathrm{max}}$, the atmosphere floor density $\rho_{\mathrm{atm}}$, number of grid points per refinement level, and the finest grid spacing employed ($\Delta x$).}
    \begin{tabular}{c c c c c c c c c c c}
    \hline\hline
         Run & $M_{\bullet}\,[M_\odot]$ &  $r_{\mathrm{H}}\,[\mathrm{km}] $ & $r_{\mathrm{ISCO}}\,[\mathrm{km}] $ & $r_{\rm in}\,[\mathrm{km}] $ & $r_{\rm out}\,[\mathrm{km}] $ & $M_{\mathrm{disk},0}\,[M_{\odot}] $ & $\rho_{\mathrm{max}}\,[{\rm g\ cm^{-3}}]$& $\rho_{\mathrm{atm}}[{\rm g\ cm^{-3}}]$ & \# cells & $\Delta x\,[\mathrm{km}]$\\ [0.5ex] 
         \hline
         M80 & 80 & 211 & 356 & 472 & 1181 & 6.7 & 2.3 $\times 10^{10}$ & 2.0 $\times 10^{-3}$ & $200^3$ & 20\\ 
       
         M500 & 500 & 1320 & 2225 & 3544 & 5759 &  75.1 & 2.5 $\times 10^9$ & 2.0 $\times 10^{-4}$ & $320^3$ & 74\\
     
         M1000 & 1000 & 2642 & 4452 & 6202 & 8860 & 125.4 & 1.4 $\times 10^9$ &  3.0 $\times 10^{-7}$ & $200^3$ & 148\\

         M3000 & 3000 & 7925 & 13356 & 19000 & 23000 & 481.0 & 6.4 $\times 10^8$ &  3.0 $\times 10^{-7}$ & $200^3$ & 266\\
         \hline
    \end{tabular}
    \label{tab:sims_info}
\end{table*}

%\footnote{\aag{The M3000 configuration is the extreme case scenario in this paper which is expected to ultimately exhibit a breakdown of ignition as discussed in ~Sec.~\ref{subsec:theo_validity_neutrino_cooling_rates} and hence might be addressed separately(from M80-M1000) in the following sections.}}

\subsection{Initial data and grid setup}
\label{sec:initial_data}

We consider four black holes of masses ${\rm (M80,M500,M1000,M3000)}=(80,500,1000, 3000)M_\odot$ and high dimensionless spin of $\chi_{\rm BH}=0.8$, spanning the range of typical massive collapsars from progenitor stars just above the pair instability supernova mass gap as discussed in \citet{Siegel_super-kilonovae_2022} to much more massive collapsars resulting from rotating SMSs (Sec.~\ref{sec:Motivation}). These black holes are initially endowed with a massive torus of cold baryonic matter (small constant specific entropy of $8k_{\rm B}$) with constant specific angular momentum and constant electron/proton fraction of $Y_e = 0.5$, corresponding to symmetric fractions of nucleons as approximately satisfied by the stable nuclei of the infalling stellar material. The basic properties of the initial BH--torus configurations are listed in Tab.~\ref{tab:sims_info}, and a visualization of the initial configurations is provided by Fig.~\ref{fig:initial_Setup}. In particular, the initial tori masses and their geometrical extend are chosen such that shortly upon relaxation and reaching an inflow equilibrium, the resulting accretion disks soon transition through the expected ignition threshold for weak interactions. The tori are embedded in a tenuous atmosphere of constant density and temperature ($T_{\rm atm}= 0.012\,\text{MeV}\simeq 1.4\times10^8$\,K), which are both sufficiently small to neither impact the dynamics nor the composition of the disk and its outflows. In particular, we ensure that the amount of atmospheric mass contained by the entire simulation grid is a few orders of magnitude smaller than the disk ejecta mass we expect.

The initial tori are endowed with a relatively weak, dynamically insignificant seed magnetic field, consisting of a single poloidal loop, specified by setting the magnetic vector potential to $A^r$, $A^{\theta} = 0$, and $A^{\phi} = A_{\rm b} \times \mathrm{max}(p - p_{\rm cut}, 0)$. Here, $A_{\rm b}$ sets the magnetic field strength to a maximum of $6.5\times10^{13}$\,G, $5.0\times10^{12}$\,G, $1.5\times10^{13}$\,G and $4.6\times 10^{13}$\,G in the M80, M500, M1000, and M3000 configurations, respectively. This results in a dynamically insignificant field with a maximum magnetic-to-fluid pressure ratio of $p_B / p < 1\times10^{-2}$. Furthermore, $p_{\rm cut}\approx 10^{-2}p_{\rm max}$, with $p_{\rm max}$ being the maximum initial pressure, specifies a threshold value that confines the non-zero magnetic field to well within the initial tori. This prevents magnetic fields to become buoyant and dynamically important in the tori surface layers during the early stages of the simulations.

The grid setup of each simulation consists of a Cartesian fixed mesh refinement level hierarchy with five concentric boxes centered upon the black hole. Each level has the same number of grid points (see Tab.~\ref{tab:sims_info}), and neighboring levels differ in a factor of two in resolution. Following earlier simulations \citep{siegel_magnetorotational_2013,siegel_three-dimensional_2018,de_igniting_2021}, the finest grid spacings as listed in Tab.~\ref{tab:sims_info}, $\Delta x=(20,74,148,266)$\,km for the M80, M500, M1000, and M3000 configurations, respectively, are chosen so that the MRI is well resolved in the quasi-stationary accretion regime that emerges after initial relaxation. This requires resolving the fastest-growing unstable mode of the MRI by at least about 8-10 grid points. The radial extent of the innermost refinement level, $L = ( 2000, 11840, 14800, 26600$)\,km for the M80, M500, M1000, and M3000 configurations respectively, is chosen so that it entirely contains the initial torus as well as the initial accretion disk upon relaxation of the torus configuration.

Additional runs at higher and lower resolution are performed to check numerical convergence of global physical quantities of interest. For the M80 configuration, we additionally employ $\Delta x = 10$\,km and $30$\,km in runs labelled M80d10 and M80d30, respectively. Regarding the configuration M500, we also run with $\Delta x = 60$\,km and $96$\,km (runs labeled M500d60 and M500d96, respectively). Similarly, for M1000 we conduct additional runs at $\Delta x = 120$\,km and $188$\,km following the same nomenclature (referred to as M1000d120 and M1000d188, respectively). Furthermore, we investigate the sensitivity of our results to the initial seed magnetic field strengths by tuning the parameter $A_b$, starting from approximately three times higher initial field strengths. These runs are labeled M80B3, M500B3, and M1000B3 for the M80, M500, and M1000 configuration, respectively. 

%The three simulations are set up and run with five fixed mesh refinement levels in a concentric Cartesian box setup and the finest resolution and (corresponding) grid size is mentioned in Table \ \ref{tab:sims_info}. The resolution increases by a factor of 2 from each finer to each coarser refinement level. The innermost refinement level grid size as presented in  Table \ \ref{tab:sims_info} is $200^3$ in M1000. For M500, the initial setup of the disk(the setup is motivated such that when the steady-state accretion sets in, the accretion rate is above the ignition threshold as described in Sec.~\ref{subsec:viscosity}) required a large refinement level boundary(and hence a box of size $320^3$) such that the refinement level doesn't cut through the disk. In M80 the setup had to be run with a large refinement level boundary but coarse resolution(leading to a box size of $130^3$) as a finer resolution(and hence a bigger box size for the same boundary) resulted in a very slow simulation(as compared to fiducial run) that early on converged to the same accretion rate as the fiducial run(Refer Sec.~\ref{subsec:accretion}). The black hole and most of the mass of the accretion disk lie completely within the innermost refinement levels at all times. The results presented in the following sections, as obtained using the diagnostic tools described in \ref{diagnostics}, are also extracted from this innermost refinement level unless otherwise stated.

\subsection{Simulation diagnostics}\label{sec:diagnostics}

We follow the nomenclature and definitions of spatial and temporal averages of physical quantities in the simulation domain as in \citet{de_igniting_2021}. In particular, we compute averages of a quantity $\chi$ weighted by the rest-mass density as
\begin{equation}
\label{eq:average_rest_mass_dens}
 \langle \chi \rangle_{\hat D} = \frac{\int \chi \hat D d^3x}{\int \hat D d^3x}.
\end{equation}
Here, $\hat D=\sqrt{\gamma}\rho W$ is the conserved rest-mass density as measured by the Eulerian observer, who moves normal to the spatial hypersurfaces of the 3+1 slicing of the spacetime. Furthermore, $\gamma=\mathrm{det}\gamma_{ij}$, where $\gamma_{ij}$ is the metric induced on the spatial hypersurfaces by the spacetime metric $g_{\mu\nu}$, and $\rho$ represents the rest-mass density. In order to characterize the physical state of the disk, it is often useful to explicitly exclude the disk corona and wind regions and thus to restrict the vertical integration only up to one scale height,
\begin{equation}
    \label{eq:rest_mass_dens_scaleht}
    \langle \chi \rangle_{\hat D, z_H} = \frac{\int^{z_H}_{-z_H} \chi \hat D \varpi dz}{\int^{z_H}_{-z_H} \hat D \varpi dz}.
\end{equation}
The local density scale height $z_H$ is obtained as
\begin{equation}
    z_H (\varpi) = \frac{\int \int_0^{2\pi} |z| \hat D \varpi d\phi dz}{\int \int_0^{2\pi}\hat D \varpi d\phi dz},
    \label{eq:scale_height}
\end{equation}
where $\varpi = \sqrt{x^2 + y^2}$ is the cylindrical coordinate radius. Additionally, in some cases it is useful to compute time averages $\langle\cdot\rangle_t$ over a time window of five consecutive time frames unless otherwise mentioned, in order to average over temporal fluctuations of physical quantities due to turbulent dynamics.

The magnetohydrodynamics (MHD)-induced turbulence in the disk introduces viscous stresses that drive accretion, outflows, and viscous spreading of the disk. These MHD stresses are fundamentally different in nature from a Shakura-Sunayev $\alpha$-viscosity \citep{shakura_black_1973}, which ensures that disks dissipate heat proportional to the local pressure (i.e.~predominantly in the midplane). Nevertheless, for purposes of comparison, one may translate the MHD stresses into an effective $\alpha$-viscosity parameter $\alpha_{\rm eff}$ that approximately characterizes the actual viscous evolution of an MHD disk. Similar to \citet{penna_shakura-sunyaev_2013} and improving upon \citet{de_igniting_2021}, we define this parameter as
\begin{equation}
\alpha_{\mathrm{eff}} = \frac{\langle\langle |t_{r\phi}|\rangle_{\hat{D},z_{H}}\rangle_t}{\langle\langle |p|  +  |\frac{b^2}{2}| \rangle_{\hat{D},z_{H}}\rangle_t},
    \label{eq:alpha_viscosity}
\end{equation}
 where
 \begin{equation}
     t_{r\phi} = \left( T_{\mu\nu}^{\rm fluid} + T_{\mu\nu}^{\rm EM}\right)e^{\mu}_{(r)}e^{\nu}_{(\phi)}, \label{eq:t_r_phi_def}
\end{equation}
with
\begin{equation}
    T_{\mu\nu}^{\rm hyd} = (\rho + \epsilon + p) h u_\mu u_\nu + p g_{\mu\nu} \label{eq:def_T_Rey}
\end{equation}
the hydrodynamic and
\begin{equation}
    T_{\mu\nu}^{\rm EM} = b^2 u_\mu u_\nu + \frac{b^2}{2}g_{\mu\nu} - b_\mu b_\nu \label{eq:def_T_Max}
\end{equation}
the electromagnetic energy-momentum tensor. Equation \eqref{eq:t_r_phi_def} represents the $r$-$\phi$ component of the stress tensor calculated in the frame comoving with the accretion flow. Here, $p$ is the fluid pressure, $\rho$ is the rest-mass density, $\epsilon$ is the specific internal energy of the fluid, and $b^2=b^\mu b_\mu$, with $b^\mu= u_\nu F^{*\mu\nu}$ being the comoving magnetic field strength, $F^{*\mu\nu}$ is the dual Faraday tensor, and $u^\mu$ denotes the 4-velocity of the fluid. The total stress in the comoving frame of the mean flow, which we approximate by $\langle t_{r\phi}\rangle_t$, consists of the Reynold's stress $t_{r\phi} = T_{\mu\nu}^{\rm hyd} e^{\mu}_{(r)}e^{\nu}_{(\phi)}$ (exclusively due to turbulent motion relative to the mean flow) and the Maxwell stress $t_{r\phi} = T_{\mu\nu}^{\rm EM} e^{\mu}_{(r)}e^{\nu}_{(\phi)}$ (due to both turbulent magnetohydrodynamic motion and coupling of fluid elements by large-scale magnetic fields). The stress $t_{r\phi}$ in the comoving frame is obtained by projecting the energy momentum tensor onto the comoving tetrad ($e^{\mu}_{(t)}$, $e^{\mu}_{(r)}$, $e^{\mu}_{(\phi)}$, $e^{\mu}_{(\theta)}$), which we derive in Appendix \ref{app:alpha_viscosity}. We generalize the tetrad computation in Boyer-Lindquist coordinates by \cite{2005_Krolik_Kerr_paper_alpha_tetrads} to an accretion flow in a general, axisymmetric, stationary background spacetime $g_{\mu\nu}$ using spherical coordinates. 

We restrict the spatial average in Eq.~\eqref{eq:alpha_viscosity} to a spatial region between cylindrical coordinate radii $\varpi \in [r_{\rm ISCO}, 3r_{\rm ISCO}]$, such that the resulting scale-height and rest-mass density averaged $\alpha_{\rm eff}$ characterizes the inner part of the accretion flow and is thus reflective of the actual measured accretion rate onto the black hole. This is for consistency, as we normalize the ignition accretion rate to its value at $r=r_{\rm ISCO}$ (Sec.~\ref{sec:theory_ign_threshold}).

\section{Simulation results}\label{sec:results}

\subsection{Relaxation of the initial data}
\label{sec:relaxation}

 \begin{figure}
     \centering
     \includegraphics[width = 0.48\textwidth]{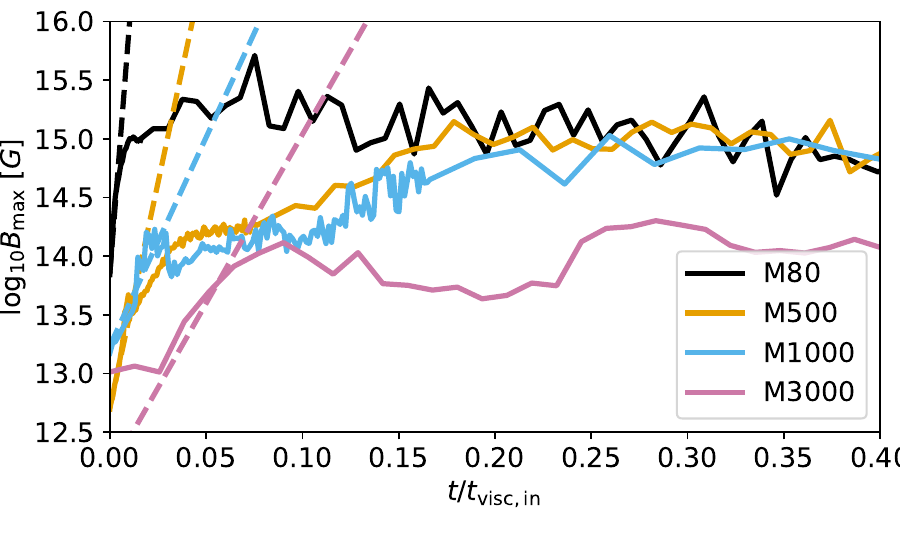}
     \caption{Maximum magnetic field strength as a function of time for the M80, M500, M1000, and M3000 accretion disk setups as a function of time in units of the initial viscous timescale (Tab.~\ref{tab:sims_alpha_info}). An initial exponential magnetic field amplification phase is evident, before the disks settle into a saturated state. The dashed lines represent exponential growth, $B_{\rm max} \propto e^{t/\tau}$, with growth times $\tau \sim 1/\Omega$, where $\Omega$ is the typical angular frequency of the respective accretion disk where dominant amplification occurs. This is the behavior expected for the linear regime of the magnetorotational instability. 
     %for M80 and $B_{\rm max} \propto e^{\Omega t/2}$ trends for M500 and M1000. The dashed lines show the exponential growth in the magnetic field due to magneto-rotational instability as expected for typical average angular velocity in the disk midplane. 
     }
     \label{fig:MRI_Bfield_amplification}
 \end{figure}
 
 \begin{figure*}[th]
 \centering
 \includegraphics[width=1.0\textwidth]{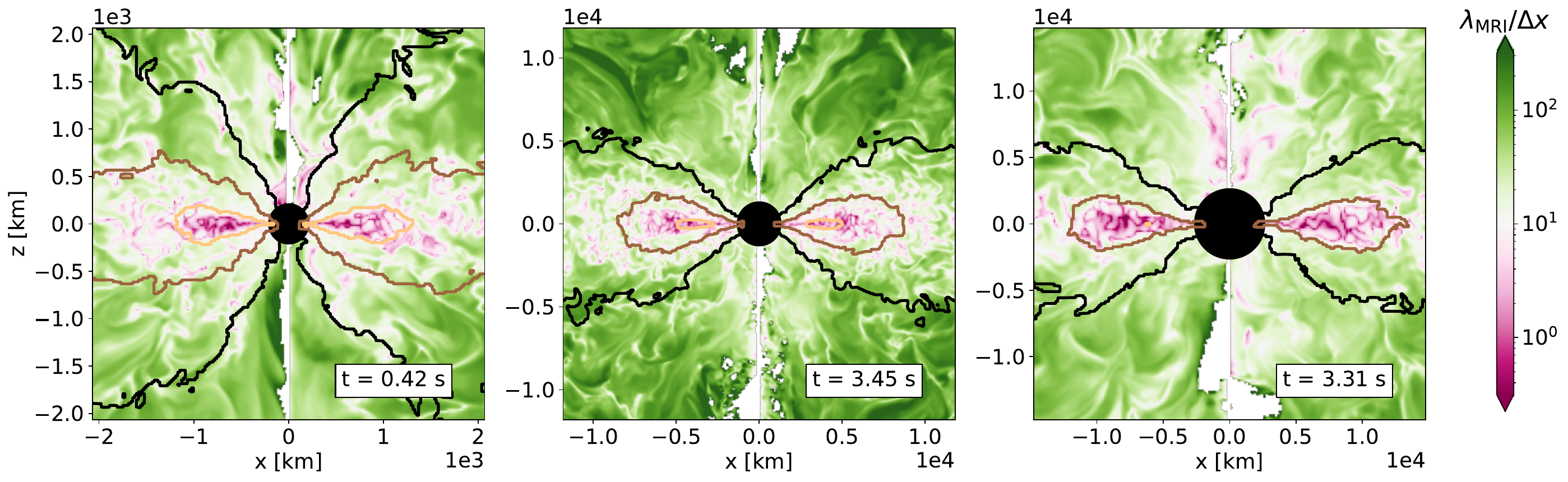}
    \caption{Snapshots of the MRI quality factor (number of grid points per wavelength $\lambda_{\rm MRI}$ of the fastest-growing unstable MRI mode) in the meridional plane after the initial torus relaxation phase, for the fiducial M80 (left), M500 (center), and M1000 (right) runs, with the fiducial M3000 setup being very similar to M1000 (not shown here). The accretion disk is indicated by contours of rest-mass density at $\rho= [10^9, 10^7, 10^5]$\,g\,cm$^{-3}$ in yellow, brown, and black, respectively. The central black holes are marked by black disks corresponding to their respective horizon radius.}
    \label{fig:lambda_MRI_resolution}
\end{figure*}
 
Upon initialization of the seed magnetic field on the differentially rotating initial torus structure (Sec.~\ref{sec:initial_data}), strong magnetic field amplification sets in as a result of the MRI, triggered in the magnetized regions with a negative radial angular velocity gradient $\mathrm{d}\Omega / \mathrm{d}\varpi < 0$ \citep{1959_Velikhov_magnetic_shearing,chandrasekhar_stability_1960,balbus_powerful_1991,balbus_instability_1998,armitage_dynamics_2011}. The ensuing relaxation phase of initial tori configurations similar to the ones considered here have been discussed in detail in previous papers (e.g., \citealt{siegel_three-dimensional_2018,Siegel_collapsars_2019,de_igniting_2021}). Here, we only comment on some essential diagnostics. 

\begin{figure*}[t]
\centering
    \includegraphics[width=1.0\textwidth]{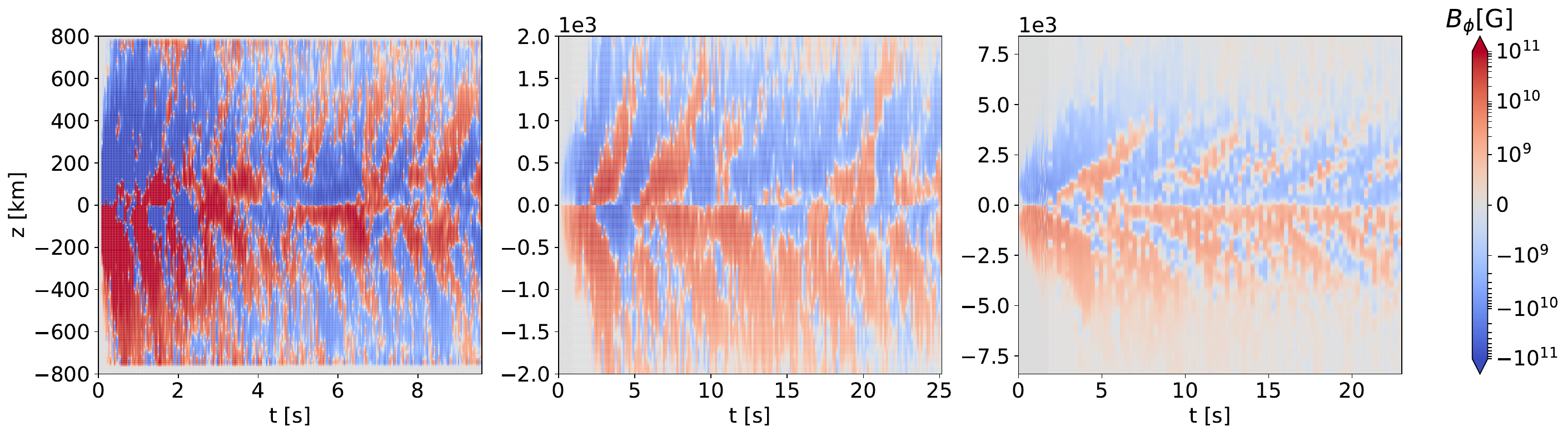}
    \caption{Space-time diagrams of the radially averaged toroidal magnetic field strength as a function of height $z$ relative to the disk midplane and time, for the M80 (left), M500 (center), and M1000 (right) configurations. Results for the fiducial M3000 setup are very similar to M1000 (not shown here). Radial averages are performed over a narrow range in radius located between $r_{\rm ISCO}$ and $3r_{\rm ISCO}$. %rin the range $r= [3-3.1],[1-1.5],[1.8-2]\ r_{\rm ISCO}$ for the M80 (left), M500 (center), and M1000 (right) configurations, respectively. 
    The apparent `butterfly signatures' indicate stationary MHD turbulence facilitated by a well-established MHD dynamo in the disk.}
    \label{fig:butterfly}
\end{figure*}

Figure \ref{fig:MRI_Bfield_amplification} shows the maximum magnetic field strength for the M80, M500, M1000, and M3000 configurations, indicating initial exponential amplification by 1--2 orders of magnitude until saturation, with an expected growth time $\tau \sim 1/\Omega$ for the linear regime of the MRI. Additional comparison runs (Sec.~\ref{sec:initial_data}) indicate that the amplification rate as well as the obtained saturation level are largely independent of both the employed resolution and the initial magnetic field strength (Fig.~\ref{fig:MRI_Bfield_amplification}). With this converged saturation level, memory of the initial data is removed, and the MRI-induced emergence of quasi-stationary MHD turbulence starts to self-consistently drive angular momentum transport and thus accretion in the resulting disk. The tori configurations quickly relax into neutrino-cooled, geometrically thin accretion disks with finite radial extend. Figure \ref{fig:lambda_MRI_resolution} reports the number of grid points per fastest growing MRI wavelength $\lambda_{\rm MRI}$, indicating that the instability is well resolved by our fiducial resolution with typically at least 8--10 grid points. Following previous work \citep{siegel_magnetorotational_2013,siegel_three-dimensional_2018}, this is sufficient to maintain a self-consistent MHD turbulent state. Indeed, upon relaxation we find that quasi-stationary butterfly signatures in the toroidal field emerge, which indicate a well-sustained MHD dynamo in our disks (Sec.~\ref{sec:accretion_tvisc}; Fig.~\ref{fig:butterfly}).

\subsection{Accretion rates and viscous evolution}
\label{sec:accretion_tvisc}

\begin{figure}[t]
    \includegraphics[width = 0.49\textwidth]{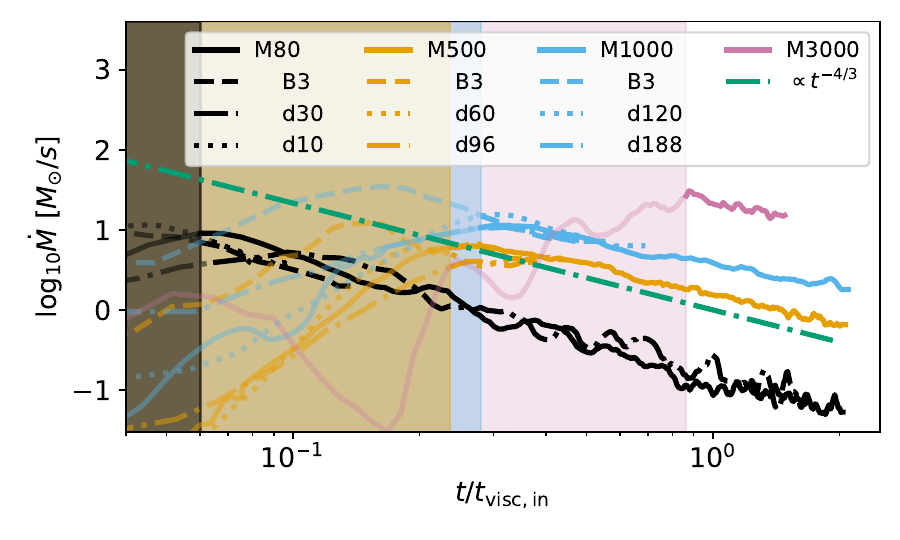}
    \includegraphics[width = 0.49\textwidth]{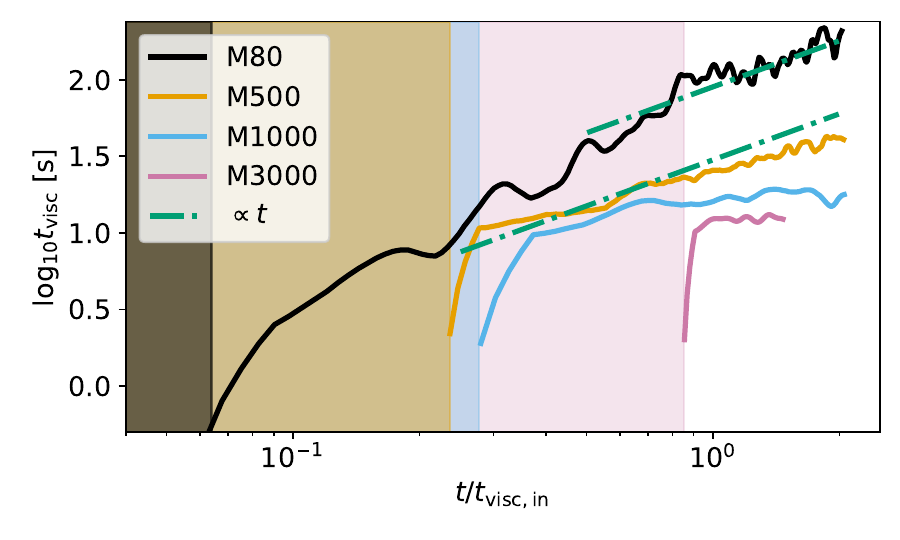}
    \caption{Top:   Accretion rate onto the black hole for the fiducial M80, M500, M1000, and M3000 accretion setups as a function of time in units of the initial viscous timescale. The shaded regions in corresponding colours mark the initial relaxation phases toward approximate inflow equilibrium. Results from additional runs with higher and lower resolution and higher initial magnetic field strength approach the same (within diagnostic limits) accretion rates upon relaxation, indicating that the fiducial setups represent approximately converged accretion regimes (see Sec.~\ref{sec:initial_data} for details). Bottom: viscous timescale as a function of time as computed from Eq.~\eqref{eq:tvsic} for the fiducial runs. The ideal late-time scalings for advective disks (ignoring disk winds) of $\dot{M}\propto t^{-4/3}$ and $t_{\rm visc}\propto t$ are also shown as dot-dashed lines. }
    \label{fig:accretion_tvisc}
\end{figure}

Upon relaxation of the initial data (Sec.~\ref{sec:relaxation}), all four accretion disk setups approach a quasi-stationary state in approximate inflow equilibrium for the inner parts of the disk. This is indicated in the top panel of Fig.~\ref{fig:accretion_tvisc} by the accretion rate $\dot M_{\rm acc}$ gradually reaching a maximum. We define the time $t_{\rm in}$ (Tab.~\ref{tab:sims_alpha_info}) at which this maximum is attained as the actual initial time of the configuration and consider the corresponding snapshot of the system as the actual initial data. The relaxation phase, which we discard from all further analysis, is indicated as a shaded region in the corresponding color in Fig.~\ref{fig:accretion_tvisc} and following figures.

Accretion rates are measured by integrating the infalling rest-mass flux through a spherical coordinate surface at radius $r=265, 1765 , 3675, 11075$ km, for the M80, M500, M1000, and M3000 setups (Tab.~\ref{tab:sims_info}), respectively. We performed resolution studies at higher resolution and additional runs with higher initial magnetic field strengths (see Sec.~\ref{sec:initial_data} for details), all of which approach the accretion rates of the respective fiducial M80, M500, or M1000 configurations (Fig.~\ref{fig:accretion_tvisc}, top panel). This indicates that the obtained accretion regimes of the fiducial setups are approximately converged upon relaxation and thus well defined. This is further corroborated by Fig.~\ref{fig:butterfly}, which shows well-established butterfly signatures emerging in the toroidal field after relaxation and for the remainder of the simulation time. Such butterfly diagrams indicate quasi-stationary MHD turbulence with a well-established MHD dynamo acting in the accretion disk. We verify that the M3000 configuration has similar accretion and magnetic field properties to M80--M1000 (including the butterfly signatures and MRI quality factor of $\sim 8-10$).

All accretion disk configurations are set up to yield stationary accretion regimes upon relaxation that are somewhat above their expected ignition threshold for weak interactions (Eq.~\eqref{eq:ignition_threshold_formula_normalized_80Msun}; see Sec.~\ref{subsec:disk_composition} for more details). We evolve all configurations over sufficient time scales, typically for more than an initial viscous time (see below), such that the accretion disks transition from an efficiently neutrino-cooled and geometrically thin regime through the ignition threshold into the inefficiently cooled, advective, and geometrically thick accretion state. In the latter late-time state, ignoring angular momentum losses through disk winds, the accretion rate is expected to follow a self-similar solution that scales as \citep{metzger_time-dependent_2008}
\begin{equation}
    \dot{M} \propto t^{-4/3}. \label{eq:Mdot_asymptotic}
\end{equation}
This is the result of continuous viscous spreading of an initially compact disk with finite extend. Such finite disk sizes arise in our context as a result of typical circularization radii of collapsar fallback material \citep{Siegel_super-kilonovae_2022}, which evolve over time depending on the progenitor structure and rotation profile. Figure \ref{fig:accretion_tvisc} shows that the behavior of Eq.~\eqref{eq:Mdot_asymptotic} is indeed approached by our numerical solutions. This further indicates that the turbulent state as shown in Fig.~\ref{fig:butterfly} self-consistently regulates accretion throughout the large (viscous) timescales of the simulations and that angular momentum transport and viscous spreading of the accretion disks is correctly captured by the numerical setup.

The bottom panel of Fig.~\ref{fig:accretion_tvisc} shows the viscous timescale as a function of time $t$ for the three fiducial simulations. Ignoring mass loss through disk winds, we compute the viscous timescale of the accretion process at a given time $t$ as
\begin{equation}
    t_{\mathrm{visc}} (t) = \frac{M_{\mathrm{disk},0} -\int_{t_{\mathrm{in}}}^{t}\dot{M}_{\mathrm{acc}}(t)\,\mathrm{d}t}{\dot{M}(t)}. \label{eq:tvsic}
\end{equation}
For the late-time advective regime of these disks, one expects the disk mass to decrease as $\propto t^{-1/3}$ \citep{metzger_time-dependent_2008}, and thus the viscous time to increase as
\begin{equation}
    t_{\rm visc}(t) \propto t. \label{eq:tvisc_scaling}
\end{equation}
While ignoring mass loss through disk winds in Eq.~\eqref{eq:tvsic} is a good approximation at early times $t\sim t_{\rm in}$, it becomes less so at later epochs. The scaling \eqref{eq:tvisc_scaling} is indeed approached by our simulations after the initial relaxation phase ($t\gtrsim t_{\rm in}$; Fig.~\ref{fig:accretion_tvisc}), further corroborating the correct behavior of angular momentum transport and viscous spreading of our disks. As expected, at late times when $t$ approaches the initial viscous timescale, deviations from the simple analytic scaling are apparent in the bottom panel of Fig.~\ref{fig:accretion_tvisc}.

We consider the relaxed state of the accretion disks as the effective initial data of all further investigation and define accordingly the initial viscous timescale $t_{\rm visc,in}$ of the respective disk as the value after relaxation at $t=t_{\rm in}$. The actual initial accretion rates and viscous timescales at this time are listed in Tab.~\ref{tab:sims_alpha_info}.

 \begin{figure}
     \centering
     \includegraphics[width=0.48\textwidth]{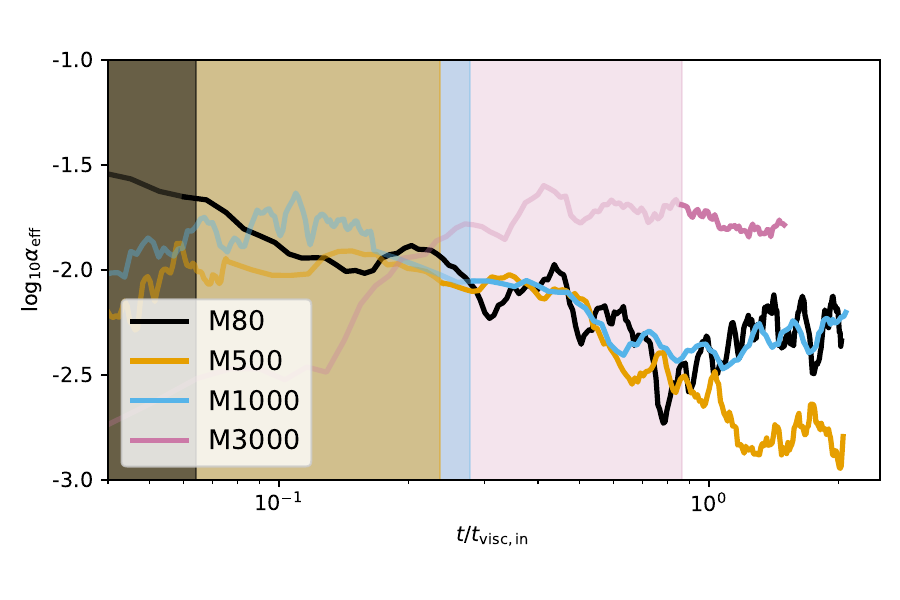}
     \caption{The effective $\alpha$-viscosity parameter as computed according to Eq.~\eqref{eq:alpha_viscosity} as a function of time for the fiducial M80, M500, M1000, and M3000 disk configurations. The shaded regions in corresponding colors mark the initial relaxation phases toward approximate inflow equilibrium. }
    \label{fig:alpha_visc}
 \end{figure}

 \begin{table*}[t]
     \caption{For the fiducial simulation runs from left to right: mass of the central black-hole ($M_{\bullet}$), initial spatially averaged effective alpha viscosity $\alpha_{\rm eff}$ (Eq.~\eqref{eq:alpha_viscosity}), corresponding ignition accretion rate ($\dot{M}_{\mathrm{ign,pred}}$) predicted by Eq.~\eqref{eq:ignition_threshold_formula_normalized_80Msun}, measured ignition accretion rate ($\dot{M}_{\mathrm{ign,sim}}$) as defined in Sec.~\ref{subsec:disk_composition}, initial viscous time scales ($t_{\mathrm{visc,in}}$) as calculated by averaging $t_{\rm visc}$ (Eq.~\eqref{eq:tvsic}) around $t_{\rm in}$, measured initial accretion rate ($\dot{M}_{\rm in} \equiv \dot{M}(t=t_{\rm in})$) time-averaged around $t_{\rm in}$, and relaxation time $t_{\mathrm{in}}$ of the initial data, which we consider as the actual initial time for all further analysis. The M3000 configuration does not achieve ignition as discussed in Sec.~\ref{subsec:disk_composition} and hence $\alpha_{\rm eff}$ (as defined in Sec.~\ref{sec:accretion_tvisc}) and $\dot{M}_{\mathrm{ign,sim}}$ are omitted. The predicted ignition accretion threshold in this case is derived for $\alpha_{\rm eff}= 1.6^{+0.4}_{-0.2}\times10^{-2}$ as extracted around $t=t_{\mathrm{in}}$. }
    \centering
    \begin{tabular}{c c c c c c c c} 
         \hline\hline
         ID & $M_{\bullet}$ [$M_\odot$] &   $\alpha_{\mathrm{eff}}$ & $\dot{M}_{\mathrm{ign, pred}}$ [$M_\odot\,\text{s}^{-1}$]& $\dot{M}_{\mathrm{ign, sim}}$ [$M_\odot\,\text{s}^{-1}$] &  $t_{\mathrm{visc,in}}$ [s] & $\dot{M}_{\mathrm{in}} $ [$M_\odot\,\text{s}^{-1}$]  & $t_{\mathrm{in}}$ [s]\\ [0.5ex] 
         \hline
         M80 & 80 & $0.9_{-0.6}^{+1.2} \times 10^{-2}$ & $0.15_{-0.1}^{+0.8}$ & $0.15_{-0.1}^{+0.5}$  & 4.7 & 9.0 & 0.3\\ 
         %\hline
         M500 & 500 & $0.7_{-0.4}^{+0.4} \times 10^{-2}$  & $1.1_{-1}^{+2}$  & $1.7_{-1}^{+3}$  & 12.3 & 6.0 & 2.9\\
         %\hline
         M1000 & 1000 & $0.9_{-0.3}^{+0.2}\times 10^{-2}$  & $4.2_{-3}^{+4}$ & $6.2_{-3}^{+5}$  & 10.8 & 10.7 & 3.0\\
         %\hline
         M3000 & 3000 & --  & $63_{-37}^{+49}$ & -- & 8.8 & 26.6 & 7.6\\
         \hline
    \end{tabular}
    \label{tab:sims_alpha_info}
\end{table*}

Figure \ref{fig:alpha_visc} shows the effective $\alpha$-viscosity $\alpha_{\rm eff}$ emerging from MHD turbulence in the inner part of the accretion disk. It drives the aforementioned accretion process and viscous spreading of the disk. Despite the large difference in black-hole masses and physical disk sizes, $\alpha_{\rm eff}$ is similar among the different configurations and tends to decrease slightly over time. Such a slight secular trend emerges as the disks viscously spread and transition through the ignition threshold into an advective, radiatively inefficient regime, in which the disks gradually become geometrically thick. The value of $\alpha_{\rm eff}$ for the relaxed M3000 configuration around $t = t_{\rm in}$ is similar to the corresponding values of M80--M1000 upon relaxation. 
%\aag{The decrease in $\alpha_{\rm eff}$ in the case of M500 is quite significant as being a massive disk(as compared to M80 and previous works) that is evolved almost $\sim 1.2$ viscous time steps (almost double that of M1000), the M500 disk is uniquely positioned to explore long term behaviour of massive accretion disks. Hence it undergoes significant changes in disk mass, size and structure which causes an eventual decrease in the comoving magnetic field which is the dominant contribution to effective alpha viscosity. }

We report the initial value of $\alpha_{\rm eff}$ at the onset of transition through the ignition threshold in Tab.~\ref{tab:sims_alpha_info}. To this end, $\alpha_{\rm eff}$ is averaged over the time frame $t_{\rm in}<t<t_{\rm ign}/2$ ($t_{\rm ign}$ being the time when $\eta_e \approx 0.5$ as described in Sec.~\ref{subsec:disk_composition}). Furthermore, using Eq.~\eqref{eq:ignition_threshold_formula_normalized_80Msun}, we calculate estimates for the ignition threshold $\dot{M}_{\rm ign}$ and list them in Tab.~\ref{tab:sims_alpha_info}. 
%The average value of $\alpha_{\rm eff}$ after relaxation and before ignition($t_{in}<t<t_{\rm ign}/2$; where $t_{\rm ign}$ being the time when $\eta_e \approx 0.5$ as described in Sec.~\ref{subsec:disk_composition}) is computed from the simulations and given in Tab.~\ref{tab:sims_alpha_info}. We use these values to estimate the expected ignition threshold $\dot{M}_{\rm ign}$ according to Eq.~\eqref{ignition_threshold_sim_scaling} as listed in Tab.~\ref{tab:sims_alpha_info}. Hence to minimise the variation in alpha-viscosity parameter value, we choose the time around $t_{\rm in}$ to calculate the average such that the disk has not entered the ignition threshold in which it goes significant structural and thermodynamically changes and the extraction of $\alpha_{\rm eff}$ becomes erroneous and fluctuating.
Due to spatial-temporal fluctuations, we estimate an uncertainty of $\alpha_{\rm eff}$ by taking the maximum and minimum value over $t_{\rm in}<t<t_{\rm ign}/2$, which characterizes the effective viscous evolution as the disks start traversing the ignition threshold (Sec.~\ref{subsec:disk_composition}). The uncertainty of the predicted value of $\dot{M}_{\rm ign}$ arises from the uncertainty in $\alpha_{\rm eff}$ as well as from the uncertainty in absolute calibration of the scaling relation itself (Eq.~\eqref{eq:ignition_threshold_formula_normalized_80Msun}), which we calibrate to the M80 case. The actual accretion rates are a factor of a few higher than the expected ignition threshold, such that all disks are indeed expected to initially reside in the neutrino-cooled state. Due to absence of ignition in the M3000 setup (Sec.~\ref{subsec:disk_composition}), $t_{\rm ign}$ is undefined and hence we do not report a value of $\alpha_{\rm eff}$ for this configuration. The expected ignition threshold in this case is calculated using $\alpha_{\rm eff}= 1.6^{+0.4}_{-0.2}\times10^{-2}$ as extracted at $t=t_{\mathrm{in}}$ with the error bars calculated from the uncertainties of normalization and $\alpha$-viscosity.

\subsection{Disk composition and transition through the ignition threshold}
\label{subsec:disk_composition}

\begin{figure*}[t]
\centering
    \includegraphics[width = 1.0\textwidth]{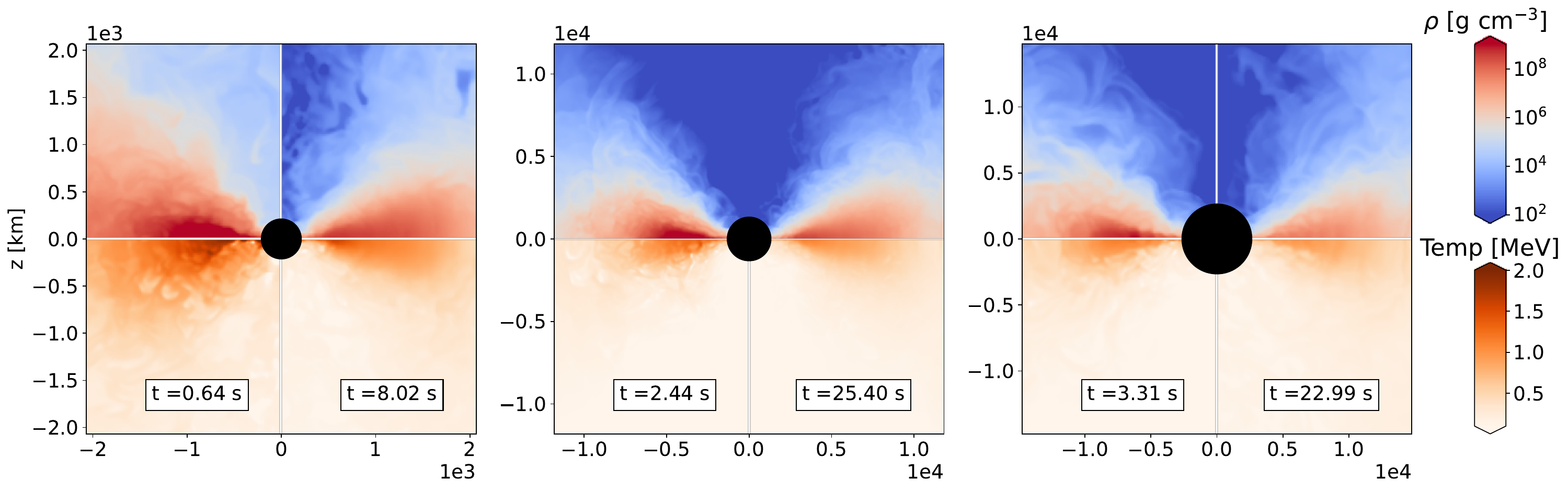}\\
    
    \includegraphics[width = 1.0\textwidth]{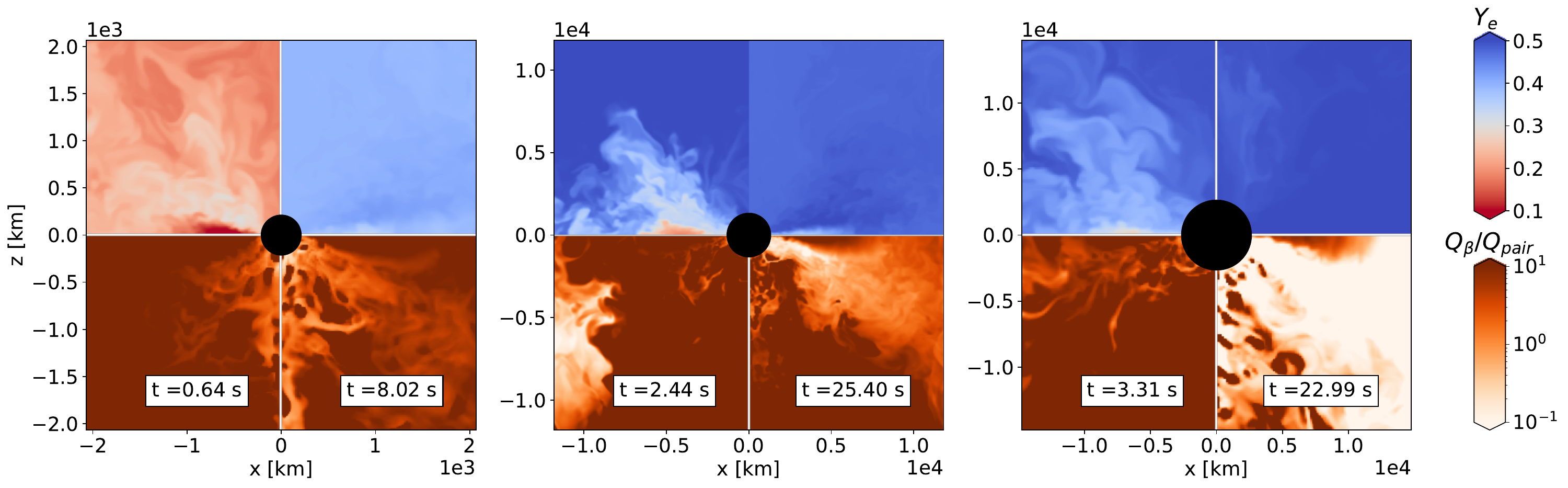}
\caption{Sections through the accretion disks along the meridional $x$-$z$ plane, comparing quantities of interest of the fiducial M80 (left), M500 (center), and M1000 (right) configurations at the initial neutrino-cooled state (left half planes) and after about a viscous timescale when they transition into the advective, geometrically thick state (right half planes). Top: rest-mass density $\rho$ (upper half-planes) and temperature $T$ (lower half-planes). Bottom: electron/proton fraction $Y_e$ (upper half-planes) and ratio of total neutrino cooling rates via $\beta$-reactions vs.~$e^\pm$-annihilation (lower half-planes). The central black hole is represented by a black disk extending up to the black-hole horizon. The relaxed M3000 configuration resides below the ignition threshold (i.e.~only in the advective, geometrically thick state) and hence is omitted here. %The density and temperatures in the disk midplane for all three simulations are of the order $>\sim 10^7$ gm/cc and 10 MeV respectively. Note that the distance scales of the three boxes are different.
}
\label{fig:rho_temperature_dual_var_image}
\end{figure*}

The fiducial M80, M500, and M1000 accretion-disk configurations start in an efficiently neutrino-cooled, geometrically thin state after relaxation at $t=t_{\rm in}$ (as expected from Tab.~\ref{tab:sims_alpha_info}) and transition through the ignition threshold within an initial viscous timescale into an advective, geometrically thick state. Figure \ref{fig:rho_temperature_dual_var_image} compares several quantities of interest in meridional sections through the accretion disks between the initial and end state of this transition. The qualitative change in scale height between these regimes is evident from the density sections. 

\begin{figure}[t]
     \centering
     \includegraphics[width = 0.48\textwidth]{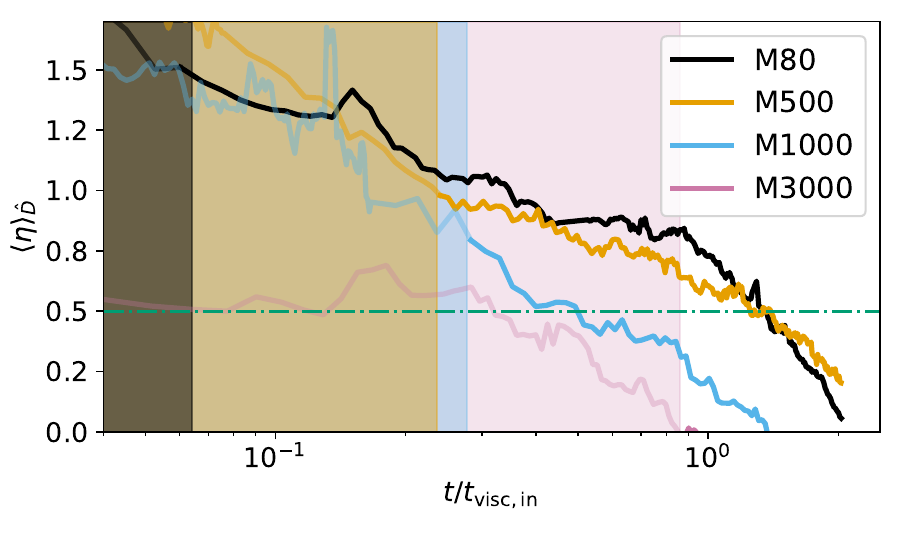} 
     \includegraphics[width = 0.48\textwidth]{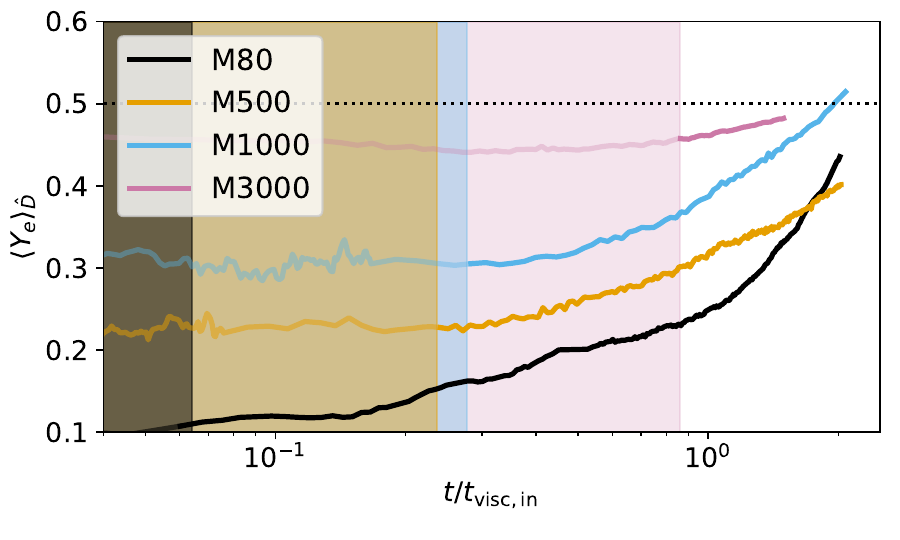}
     \caption{Rest-mass density averaged, normalized electron degeneracy $\eta$ (top) and electron/proton fraction $Y_e$ (bottom) of the fiducial M80, M500, M1000, and M3000 accretion disks as a function of time. In the top panel, $\eta = 0.5$ is shown with a horizontal dotted-dashed line to approximately indicate the transition through the ignition threshold. Residual neutronization ($Y_e\lesssim 0.5)$ in the M3000 run is due to the initial relaxation phase. Averages are computed over a spatial range of $[r_{\rm ISCO}, 3 r_{\rm ISCO}]$. The shaded regions in corresponding colours mark the initial relaxation phases toward approximate inflow equilibrium.
     }
     \label{fig:eta_Ye}
 \end{figure}

At $\dot{M}>\dot{M}_{\rm ign}$, the midplane densities are sufficiently large for electrons to be degenerate. Once the normalized chemical potential $\eta_e = \mu_e / k_{\rm B}T > 1$, where $\mu_e$ denotes the chemical potential of electrons and $k_{\rm B}$ the Boltzmann constant, a self-regulation mechanism within the disk sets in, which drives the accreting plasma to $\eta_e\sim 1$ \citep{chen_neutrino-cooled_2007,siegel_three-dimensional_2017,siegel_three-dimensional_2018}. This can be seen in the top panel of Fig.~\ref{fig:eta_Ye}, which reports $\eta_e\approx 0.9-1.5$ for the M80--M1000 disks after the relaxation phase.

%The estimated value(and error bars) is obtained from Eq.~\eqref{eq:ye_ignition_threshold} using corresponding averages(and the corresponding maximum and minimums) for $\eta_e$ and temperature. Due to significant changes during evolution in the disk structure, temperature and degeneracy change significantly causing large error bars in the $Y_e$ estimate.}

As a result of electron degeneracy, pair creation is suppressed due to Fermi blocking, which, in turn suppresses positron capture onto neutrons relative to electron capture onto protons. As a result, the accretion flow neutronizes irrespective of the initial composition of the plasma \citep{Siegel_collapsars_2019}. For weakly degenerate conditions, $\eta_e \lesssim 1$, and hot accretion flows $\theta \equiv k_{\rm B}T/m_e c^2 > Q=(m_n-m_p)/ m_e \approx 2.531$ (i.e., $T > 1.3$\,MeV), the resulting proton fraction $Y_e$ can be analytically estimated as (\citealt{beloborodov_nuclear_2003,siegel_three-dimensional_2018})
\begin{equation}
    Y_e = 0.5 + 0.487\left(\frac{1.2655}{\theta} - \eta\right).
    \label{eq:ye_ignition_threshold}
\end{equation}
Despite the fact that the temperature constraint $\theta > Q$ is marginally violated for the M500 and M1000 configurations (see Sec.~\ref{subsec:results_temperature}, Fig.~\ref{fig:Ye_trisim}), we find these analytic predictions to be in rough agreement with the average $Y_e$ extracted from our accretion disks as they transition through ignition. % I would suggest to keep this statement, otherwise we underminde our analysis regarding the scaling analysis further down %\aag{However, due to low disk temperatures of $T\lesssiM80.3$ MeV(see Sec.~\ref{subsec:results_temperature}, Fig.~\ref{fig:Ye_trisim}), we find these analytical predictions do not align with the inferred $Y_e$ from the simulations.}

%In particular, the disk midplane temperature stays at $T\gtrsim 1\,$ MeV for all disks for times of interest (Fig.~\ref{fig:rho_temperature_dual_var_image}, top row). 

\begin{figure}[t]
    \centering
\includegraphics[width=1.0\linewidth]{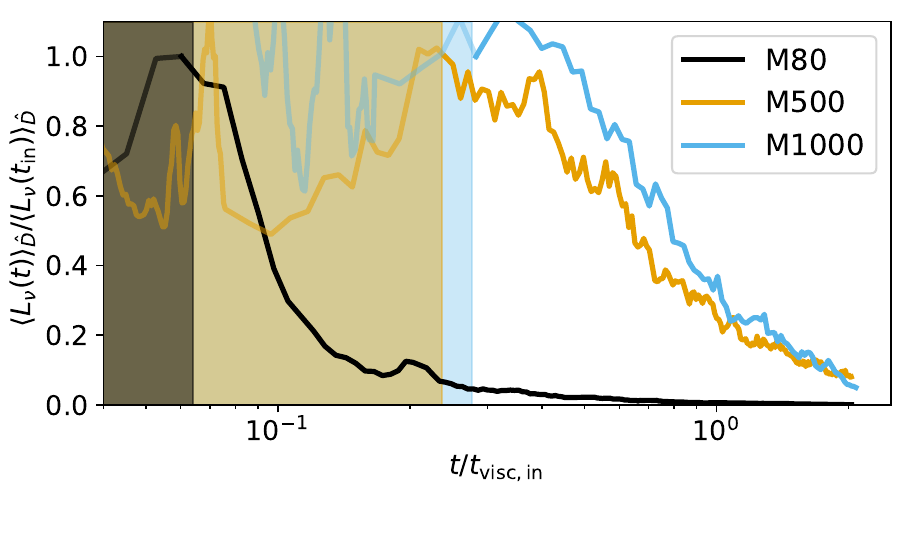}
    \caption{The total neutrino cooling luminosity for the fiducial M80, M500, and M1000 accretion disks as a function of time in units of the initial viscous timescale, normalized to the value at $t=t_{\rm in}$ (just after the initial relaxation period corresponding to the shaded areas). A rapid decrease over one order of magnitude from an initial `plateau' is evident as the disks transition through the ignition threshold, becoming radiatively inefficient. The shaded regions in corresponding colours mark the initial relaxation phases toward approximate inflow equilibrium. The relaxed M3000 configuration (not shown) does not reach ignition and thus neutrino emission is heavily suppressed.}
    \label{fig:nu_cooling_rates}
\end{figure}

As the accretion disks viscously spread (Sec.~\ref{sec:accretion_tvisc}), the midplane density decreases and so does the electron degeneracy $\eta_e$. At a slowly varying disk temperature (as observed in the M80, M500, and M1000 configurations), this implies an increase in $Y_e$ according to Eq.~\eqref{eq:ye_ignition_threshold}. Figure \ref{fig:eta_Ye} shows a steepening decline in $\eta_e$ and a corresponding increase in $Y_e$ as the disks evolve over an initial viscous timescale. The disks thereby transition from an efficiently neutrino-cooled, geometrically thin to an inefficiently cooled, geometrically thick state, as illustrated by Figs.~\ref{fig:rho_temperature_dual_var_image} and \ref{fig:nu_cooling_rates}. In particular, the neutrino luminosities start at a respective `plateau' just after relaxation and then quickly drop by an order of magnitude as the disks approach and transition through the ignition threshold (Fig.~\ref{fig:nu_cooling_rates}). We (somewhat arbitrarily) define the ignition threshold as the average physical state of the disk over the time range in which $\langle\eta_e\rangle_{\hat D}\in [0.1,0.9] $. We define the time $t=t_{\rm ign}$ when $\langle\eta_e\rangle_{\hat D} \approx 0.5$, indicated by a horizontal dashed line in the top panel of Fig.~\ref{fig:eta_Ye}. We assign uncertainties to physical quantities of interest at the ignition threshold by extracting the maximum and minimum over the same time window.

The M3000 configuration leads to $\langle\eta_e\rangle_{\hat{D}} < 0.1$ upon relaxation (Fig.~\ref{fig:eta_Ye}, top panel). Therefore, the M3000 configuration does not achieve ignition at the expected accretion rate (Tab.~\ref{tab:sims_alpha_info}) and, accordingly, does not show significant neutronization of the accretion flow (Fig.~\ref{fig:eta_Ye}, bottom panel). The residual neutronization is due to the relaxation phase, in which $\langle\eta_e\rangle_{\hat{D}} \lesssim 0.5$ is reached owing to the compact nature of the initial torus. The gradual increase in $\langle Y_e\rangle_{\hat{D}}$ after relaxation toward $\langle\eta_e\rangle_{\hat{D}} = 0.5$ evident in Fig.~\ref{fig:eta_Ye} shows that residual neutrino cooling is still active, but neutronization is prevented due to vanishing electron degeneracy. The measured accretion rate of the M3000 run thus serves as a lower limit on the ignition threshold. Ignition may still be achieved at (much) higher accretion rates of $\gtrsim 100\ M_{\odot}/s$, although these become increasingly less plausible from an astrophysical point of view. Moreover, such high accretion rates would likely render the accretion disk gravitationally unstable \citep{gammie_nonlinear_2001}, which impacts the disk structure, accretion rate, and, likely, ignition.

%The bottom panel of Fig.~\ref{fig:nu_cooling_pressure_ratios} shows the average ratio of radiation pressure to total pressure (baryons, radiation, and magnetic pressure) as the M80, M500, and M1000 disks transition through the ignition threshold. This verifies a second important assumption in deriving the approximate scaling relation for $\dot{M}_{\rm ign}$, namely that radiation pressure and pressure due to relativistic $e^\pm$, both of which are characterized by $p\propto T^4$, dominate over baryon pressure. While 
%across the entire mass range of black holes considered here above ignition threshold and as the disk accretion rate transitions through ignition threshold, radiation and electron pressure become dominant.

\subsection{Scaling of the ignition threshold}
\label{sec:results_scaling_ignition_threshold}

\begin{figure}[tb]
     \centering
     \includegraphics[width=0.47\textwidth]{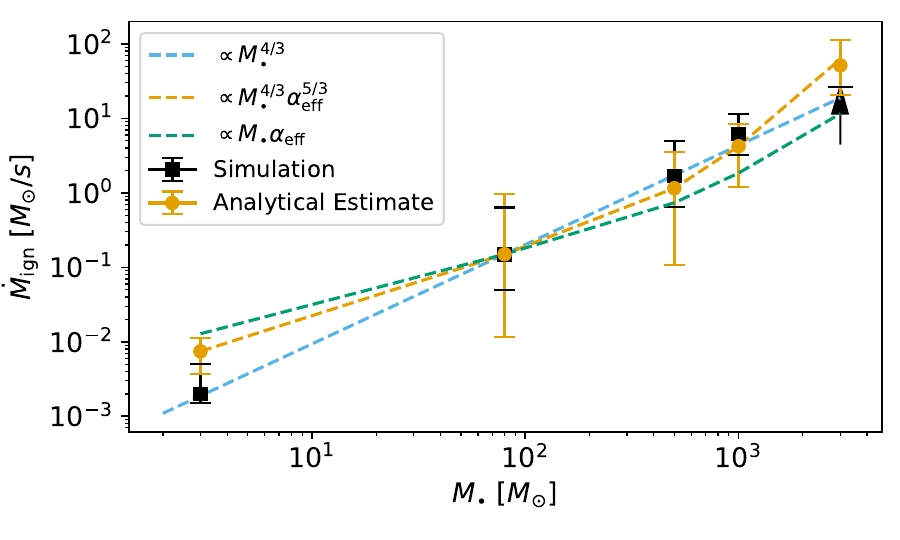}
     %\hspace*{-0.6cm}\includegraphics[width=0.49\textwidth]{figures/no_alpha_scaling.pdf}
     \caption{The ignition accretion rate extracted from the M80, M500, and M1000 simulations as an average over the time frame in which $\eta_e \in[0.1,0.9]$ (black squares), with uncertainties 
     representing the range of accretion rates in this time window. The simulation data for $M_\bullet = 3M_\odot$ are taken from \citet{de_igniting_2021}. Orange dots represent the corresponding analytical estimates (Eq.~\eqref{eq:ignition_threshold_formula_normalized_80Msun}), with error bars arising from $\alpha_{\rm eff}$ (Tab.~\ref{tab:sims_alpha_info}, Sec.~\ref{sec:accretion_tvisc}) and calibration to the M80 setup. Comparisons to a pure $\propto M_{\bullet}^{4/3}$ scaling and to a scaling $\propto M_{\bullet}\alpha_{\rm eff}$ appropriate for the neutrino opaque threshold with dominant baryon pressure (Eq.~\eqref{eq:Mdot_opaque}) are also shown. The relaxed M3000 accretion disk configuration resides below the ignition threshold and thus provides a lower limit (black arrow) on the ignition accretion rate.}
     \label{fig:Mdot_ign_comparison}
 \end{figure}

Figure \ref{fig:Mdot_ign_comparison} compares the ignition threshold accretion rates from our fiducial M80, M500, and M1000 disk simulations, extracted when the disks reach $\langle \eta_e \rangle_{\hat D} = 0.5$ (averaged over the time window $\langle\eta_e\rangle_{\hat D}\in [0.1,0.9]$), to corresponding analytic estimates based on Eq.~\eqref{eq:ignition_threshold_formula_normalized_80Msun}. The estimates show remarkable agreement with the actual extracted values of $\dot{M}_{\rm ign}$ within error bars across three orders of magnitude in black-hole mass. In the suite of simulations presented here, we have added results at $M_\bullet = 3M_\odot$ from \citet{de_igniting_2021}. Significant uncertainties in the analytic estimates arise from turbulent fluctuations of the effective alpha viscosity $\alpha_{\rm eff}$ (Tab.~\ref{tab:sims_alpha_info}, Sec.~\ref{sec:accretion_tvisc}). This highlights the conceptual and practical problems as well as limitations of reducing viscosity due to complex spatio-temporal, three-dimensional magnetohydrodynamic phenomena to a single effective constant. In particular, agreement between the analytical estimate of $\dot{M}_{\rm ign}$ and the simulation-extracted value for $M_\bullet = 3M_\odot$ is of the least quality, which we attribute to a different numerical extraction of $\alpha_{\rm eff} = 0.02_{-0.01}^{+0.01}$ in \citet{de_igniting_2021}. Despite these uncertainties, our simulation results across the large black-hole mass range of $3-1000\,M_\odot$ are sufficiently accurate to favor a scaling law according to Eq.~\eqref{eq:ignition_threshold_formula_normalized_80Msun} over a simple scaling law $\propto M_\bullet^{4/3}$ (neglecting the dependence on $\alpha_{\rm eff}$) or over a slightly modified scaling law $\propto M_{\bullet}\alpha_{\rm eff}$, as expected for the neutrino opaque threshold assuming dominant baryon pressure (Eq.~\eqref{eq:Mdot_opaque}).

%\aag{The inferred ignition threshold is compared with the estimated values in Fig.~\ref{fig:Mdot_ign_comparison} and the two overlap very well(within error margins) across three orders of central black-hole mass. The apparent deviation between expected and inferred values arises due to the heavy fluctuations and corresponding uncertainties in the effective alpha viscosity($\alpha_{\rm eff}$) parameter(Ref Tab. ~\ref{tab:sims_alpha_info}). This highlights the limits and problems of reducing viscosity due to complex three-dimensional magneto-hydrodynamic phenomena to a single effective parameter. In the bottom panel of Fig.~\ref{fig:Mdot_ign_comparison} we have shown the ignition threshold scaling without the effective alpha-viscosity factor, which indeed shows a better fit due to reduced uncertainties from omitting $\alpha_{\rm eff}$.}

The definition of the ignition threshold adopted here is based on electron degeneracy, a criterion that can be monitored self-consistently and precisely for all simulations. However, it differs somewhat from the definition adopted in Sec.~\ref{sec:theory_ign_threshold} and from that adopted by one-dimensional, stationary disk models \citep{chen_neutrino-cooled_2007,Siegel_collapsars_2019,de_igniting_2021,Siegel_super-kilonovae_2022}, 
% add hernandez-morales_neutrino-cooled_2025 when submitted to arXiv
which typically base it on a ratio of neutrino cooling to viscous heating. Lacking a self-consistent and precise measure of viscous heating in our GRMHD simulations, electron degeneracy appears to be a more robust approach. Since significant electron degeneracy ($\eta_e\approx 0.5$) typically requires slightly denser environments compared to those at the ignition threshold in one-dimensional, viscous models, our ignition threshold is somewhat skewed toward higher accretion rates. Despite these differences, Fig.~\ref{fig:Mdot_ign_comparison} shows that the scaling we find is still in remarkable agreement with that predicted based on comparable neutrino cooling vs.~viscous heating. Even slight modifications of the scaling law are disfavored based on our simulation results.

%\aag{The definition of ignition threshold as the accretion rate regime when $\eta_e \in [1,0.1] $ is chosen due to electron degeneracy being an intrinsic(to the simulations and the hydrodynamical variables of the disk) indicator of weak interactions. However, it is arbitrary and less precise as compared to the criterion used in previous works \citep{de_igniting_2021,Siegel_super-kilonovae_2022} to derive the scaling relation(Eq.~\ref{ignition_threshold_sim_scaling}). The latter is defined as the accretion rate at which the cooling due to neutrinos is more than half of the viscous heating rate. However, the viscous heating rate is heavily spatially dependent and difficult to condense into an effective heating rate(following similar pathologies as $\alpha_{\rm eff}$) for the whole disk. This spatial dependence also implies that weak interactions in different parts of the disk are ignited at different times which served as a motivation for selecting a degeneracy band of $\eta_e \in [1,0.1] $ instead of an arbitrarily chosen precise value.}

\subsection{Break-down of ignition}
\label{sec:results_breakdown_ignition}

In the following, we analyze how well the assumptions leading to an ignition threshold and its particular scaling with black-hole mass and disk viscosity (Sec.~\ref{sec:results_scaling_ignition_threshold}) are satisfied in our simulations, and whether we see evidence for a break-down of ignition at large black-hole masses. 

\begin{figure}[t]
    \centering
      \includegraphics[width = 0.47\textwidth]{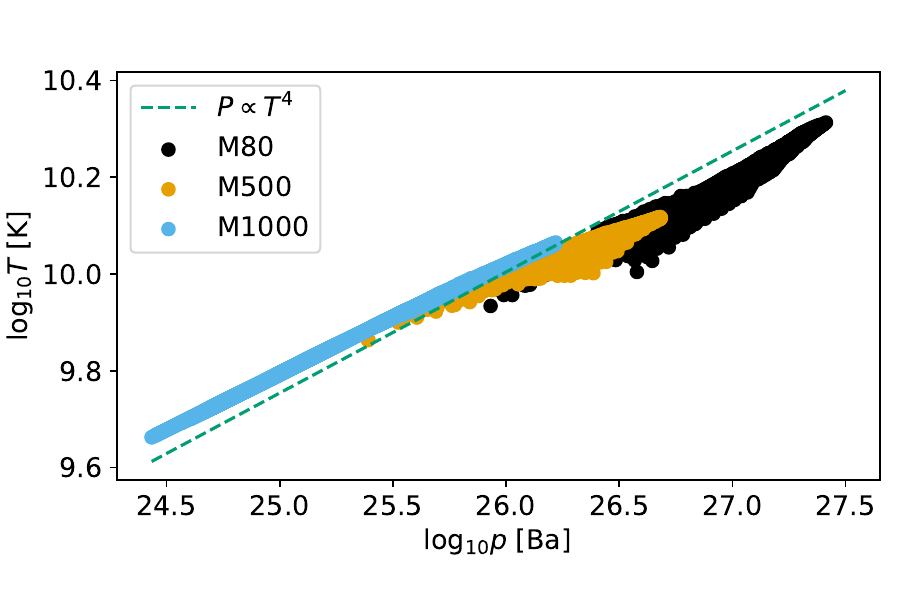}
     \caption{Total pressure as a function of temperature for grid points in the $xz$-plane within one scale height off the disk midplane and in a radial range of $[1.5r_{\rm ISCO}, 3 r_{\rm ISCO}]$ at the time when the M80, M500, and M1000 disks transition through their respective ignition threshold ($\eta\approx 0.5$). The approximate scaling $p\propto T^4$ is expected from dominant radiation pressure and relativistic $e^\pm$-pressure. The relaxed M3000 accretion disk configuration resides below the ignition threshold and is therefore omitted here.}
     \label{fig:pressure_ratios}
 \end{figure}

\subsubsection{Pressure of the accretion flow}
\label{subsec:results_pressure}

Figure~\ref{fig:pressure_ratios} shows the total pressure as a function of temperature for grid points in the $xz$-plane within one scale height of the disk midplane in the innermost part of the accretion disk at the time when the M80, M500, and M1000 disks transition through their respective ignition threshold ($\eta\approx 0.5$). This verifies a second important assumption in deriving the approximate scaling relation for $\dot{M}_{\rm ign}$, namely that the total pressure approximately follows $p\propto T^4$, which is expected from dominant radiation and relativistic $e^\pm$-pressure. The fact that the scaling approaches $p\propto T^4$ more closely with increasing black-hole mass is expected based on Eq.~\eqref{eq:Pressure_Scaling}. The numerical simulations thus corroborate the analytic estimates and a modification of the fiducial ignition scaling relation with increasing black-hole mass based on pressure contributions is not expected.

\subsubsection{Neutrino opacities and opaqueness threshold}

\begin{figure}
    \centering
    \includegraphics[width=1.0\linewidth]{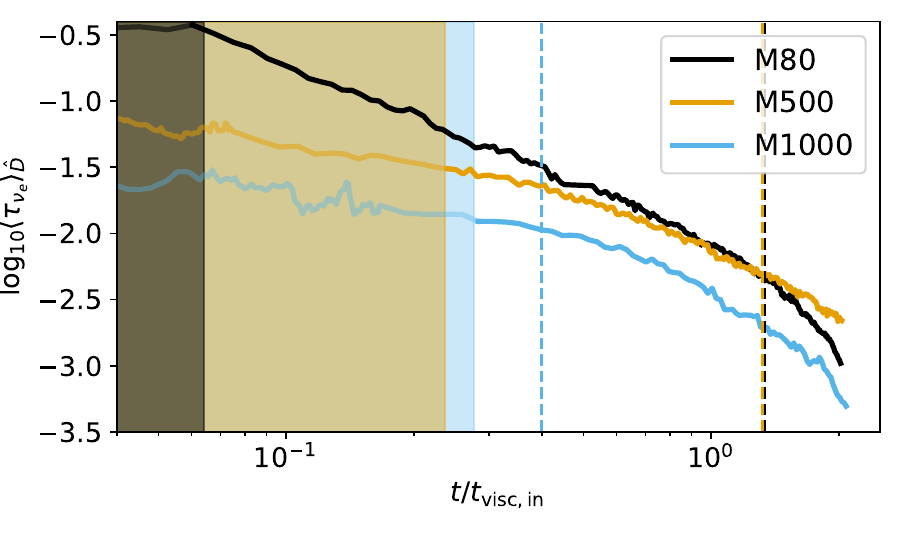}
    \caption{Average optical depth to electron neutrino number emission as a function of time for the fiducial M80, M500, M1000 disk setups. The optical depths are well below unity at the respective ignition threshold (represented by vertical dashed lines), indicating that the accretion flow is not opaque to neutrinos in this regime. The shaded regions in corresponding colors mark the initial relaxation phases toward approximate inflow equilibrium. The relaxed accretion disk of the M3000 configuration resides below the ignition threshold and hence is omitted here.}
    \label{fig:electronneutrino_number_optical_depth}
\end{figure}

Figure~\ref{fig:electronneutrino_number_optical_depth} reports the weighted average neutrino optical depth $\tau_{\nu_e}$ for electron neutrinos in the inner part of the accretion disk for the M80--M1000 fiducial simulations. The optical depths are computed on the numerical grid by a quasi-local scheme as described in \citet{siegel_three-dimensional_2018}. The optical depth remains at $\tau_{\nu_e}\ll 1$ at the ignition threshold for the M80--M1000 simulations. Thus the accretion flow clearly resides in the optically thin regime up to $M_\bullet \sim 1000 M_\odot$ as assumed in deriving the ignition threshold in Sec.~\ref{sec:theory_ign_threshold}. The disks thus do not enter a possible crossing of $\dot{M}_\nu$ and $\dot{M}_{\rm ign}$ (Sec.~\ref{subsec:theo_optically_thin}) before the break-down of standard ignition in the M3000 configuration at $M_\bullet = 3000 M_\odot$.

%lies much below one($<<1$), hence there is almost no or minimal neutrino trapping in the disk-midplane. Thus neutrinos are effectively emitted from the disk and are the dominant cooling mechanism due to their minimal interaction nature.

\subsubsection{Neutrino cooling rates}
\label{subsec:results_neutrino_cooling}

\begin{figure}[t]
    \centering
     \includegraphics[width = 0.47\textwidth]{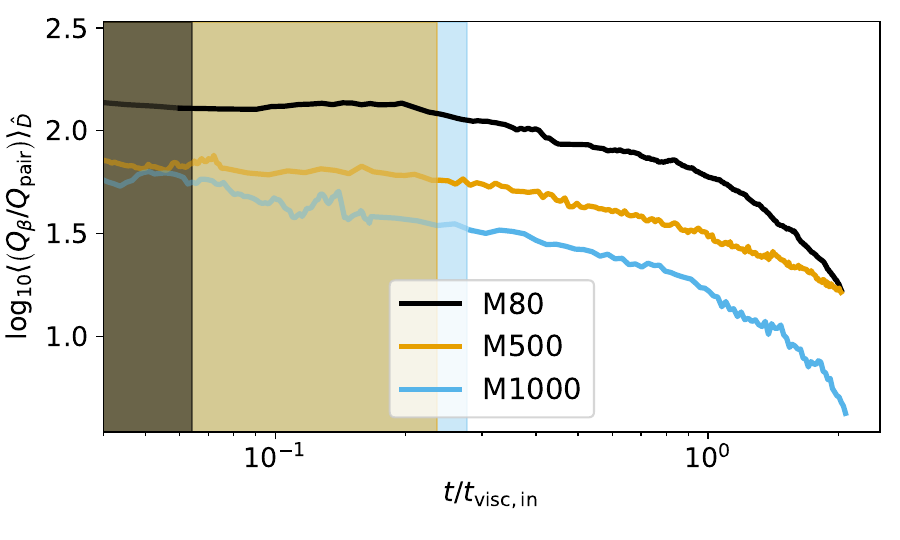}
    \caption{Average ratio of total effective neutrino cooling rates via $e^\pm$-capture (indicated as $\beta$-reactions) to $e^\pm$-pair annihilation for the M80, M500, and M1000 accretion disks as they transition through the ignition threshold.  Averages are computed over a spatial range of $[1.5r_{\rm ISCO}, 3 r_{\rm ISCO}]$. Pair annihilation remains subdominant above the ignition threshold and across the large range of black hole masses of 80--1000\,$M_\odot$ considered here, although they become increasingly relevant as the disks migrate to lower disk densities into the advective regime. The shaded regions in corresponding colours mark the initial relaxation phases toward approximate inflow equilibrium. Neutrino emission is strongly suppressed in the M3000 configuration, which is thus omitted here.}
     \label{fig:nu_cooling}
\end{figure}

As the M80, M500, and M1000 configurations transition through the ignition threshold, electron and positron capture remain by far the dominant cooling mechanism of the disks. Figure \ref{fig:nu_cooling} reports the ratio of effective net cooling rates via the $\beta$-reactions to the effective cooling rates through $e^\pm$-pair annihilation, which remains $>\!10$ throughout the transition. Spatially resolved snapshots before and after the transition are shown in Fig.~\ref{fig:rho_temperature_dual_var_image}. Dominance of $\beta$-reactions clearly correlates with the high-density midplane regions, while pair annihilation is important in the low-density regions, especially in the disk corona. The latter, however, only contributes to a minor degree to the overall cooling of the accretion flow, due to strong vertical gradients in density and temperature off the disk midplane and very sensitive scaling $\propto \rho T^9$ of the specific emissivities for pair annihilation \citep{qian_nucleosynthesis_1996,Burrows_2006_neutrino_opacities}. 

These results verify the assumption underlying the derivation of the fiducial scaling relation for $\dot{M}_{\rm ign}$ that other cooling mechanisms beyond electron and positron capture can be neglected, across the entire mass range of 80--1000\,$M_\odot$ of the black holes considered here. A modification of the expected scaling of $\dot{M}_{\rm ign}$ with $\alpha_{\rm eff}$ and $M_\bullet$ due to pair annihilation and, possibly, plasmon decay is thus expected to occur only at much higher black-hole masses than $\gtrsim 1000\,M_\odot$. However, the modified scaling $\dot{M}_{\rm ign}\propto \alpha_{\rm eff}^{9/5}M_\bullet^{6/5}$ resulting from dominant pair annihilation ($T^{9}/\rho \propto \alpha r^{2}\Omega^{3}$ in lieu of Eq.~\eqref{eq:midplane_temperature_Mign_1}) is remarkably close to that resulting from the dominance of electron and positron captures. Distinguishing these power laws from each other using three-dimensional simulations of the type presented here would be difficult diagnostic-wise.

\subsubsection{Temperature of the accretion flow and breakdown of neutronization}
\label{subsec:results_temperature}

\begin{figure}
     \centering
     \includegraphics[width = 0.47\textwidth]{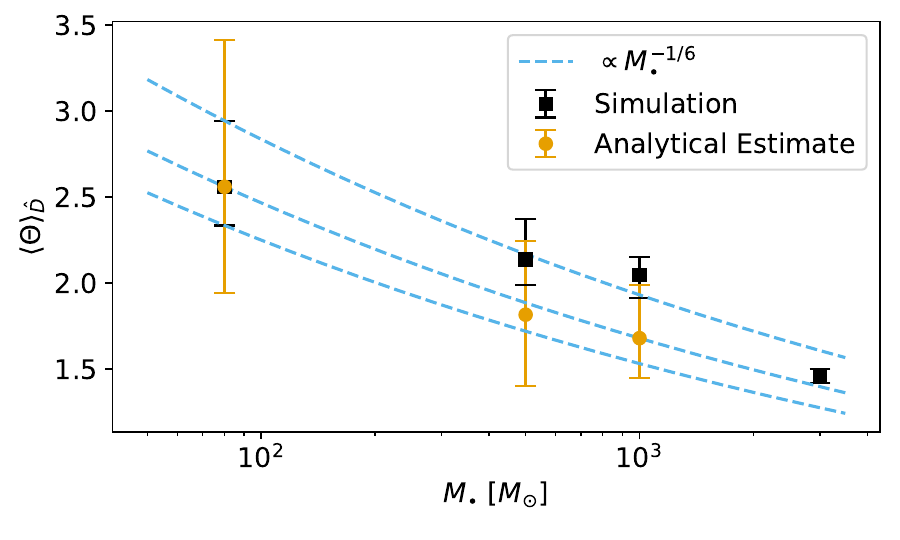}
     \includegraphics[width = 0.47\textwidth]{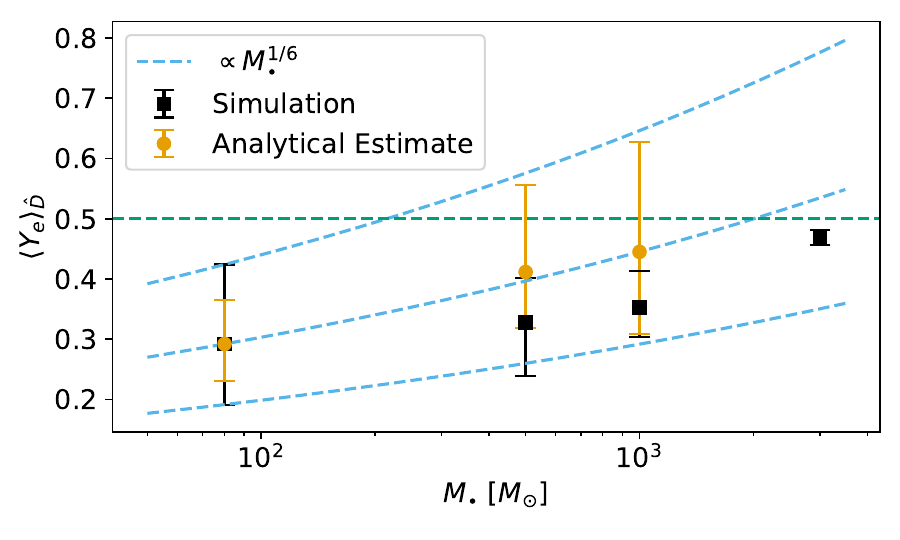}
     \caption{ 
     Average normalized temperature $\theta$ (top) and proton fraction $Y_e$ (bottom) of the fiducial M80, M500, M1000, and M3000 configurations, temporally averaged around the transition time of the ignition threshold over a time frame in which $\eta_e\in [0.1,0.9]$ and represented as black squares (for M3000, $\eta_e < 0.1$ in the relaxed state and hence an averaging window $t > t_{\rm in}$ is considered). Error bars correspond to the maximum and minimum value of the quantity attained in the temporal averaging window. Orange data points represent analytical estimates that are calculated according to Eqs.~\eqref{eq:scaling_temperature_BHmass} and \eqref{eq:scaling_Ye_BHmass}, with error bars arising from $\alpha_{\rm eff}$ (Tab.~\ref{tab:sims_alpha_info}, Sec.~\ref{sec:accretion_tvisc}; $\alpha_{\rm eff}$ is undefined for M3000 and hence the analytical estimate is omitted) and calibration errors. Also shown as dashed lines are the analytically estimated trends (without $\alpha_{\rm eff}$) according to Eqs.~\eqref{eq:scaling_temperature_BHmass} and \eqref{eq:scaling_Ye_BHmass}.}
     \label{fig:Ye_trisim}
 \end{figure}

The temperature of accretion flows at the ignition threshold for plasma with $p\propto T^4$ (see above, Sec.~\ref{subsec:results_pressure}) is expected to decrease with black-hole mass as (Eq.~\eqref{eq:temperature_scaling})
\begin{equation}
    T,\theta\big|_{\dot{M}_{\rm ign}} \propto \alpha_{\rm eff}^{1/6}M_{\bullet}^{-1/6}. \label{eq:scaling_temperature_BHmass}
\end{equation}
As illustrated in the top panel of Fig.~\ref{fig:Ye_trisim}, we recover this trend remarkably well in our simulations across the wide black-hole mass range considered here.

The mean temperature in the M500, M1000, and M3000 accretion disks is below the neutron-proton mass difference, $\theta \equiv k_{\rm B}T/m_e c^2 < Q=(m_n-m_p)/ m_e \approx 2.531$, which translates to $T < 1.3$\,MeV (Fig.~\ref{fig:Ye_trisim}, top panel). Despite the resulting onset of suppression of neutronization at these large black-hole masses, the bottom panel of Fig.~\ref{fig:Ye_trisim} shows that we still obtain (up to a constant offset) the expected scaling of the average proton fraction at the ignition threshold following from Eqs.~\eqref{eq:ye_ignition_threshold} and \eqref{eq:scaling_temperature_BHmass},
\begin{equation}
    Y_e \big|_{\dot{M}_{\rm ign}}\propto \alpha_{\rm eff}^{-1/6}M_\bullet^{1/6}. \label{eq:scaling_Ye_BHmass}
\end{equation}
This is consistent with the expectation that the transition to inefficient neutrino-cooling is gradual, thanks to the cooling rates being skewed toward the high-energy tail of $e^\pm$ in the Fermi-Dirac distribution (see Sec.~\ref{subsec:theo_validity_neutrino_cooling_rates}).
%The difference in $Y_e$ between the M80 and M500 configurations at ignition is mainly due to higher temperatures in the M80 case, while the difference in $Y_e$ between the M500 and M1000 disks is due to higher degeneracy $\eta_e$ in the M500 case (Fig.~\ref{fig:eta_Ye}, top panel).

With an average proton fraction of $Y_e\approx 0.3$ in the disk (Fig.~\ref{fig:Ye_trisim}) as well as at the base of the disk winds (Fig.~\ref{fig:rho_temperature_dual_var_image}) at the ignition threshold, the M1000 configuration is still expected to lead to r-process nucleosynthesis in the outflowing plasma. Peaking at $Y_e > 0.25$, the composition of both the M500 and M1000 disk outflows will predominantly synthesize light r-process elements, given the expansion speeds of $v\approx 0.1 c$ and entropies of $s\gtrsim 20 k_{\rm B}$ per baryon of the disk outflows (\citealt{lippuner_r-process_2015}; see also \citealt{siegel_three-dimensional_2018,Siegel_collapsars_2019}). For the M3000 configuration, not only is the disk temperature well below the neutron-proton mass difference, there is also little to no neutronization anymore ($Y_e \simeq 0.5$), as expected from Eq.~\eqref{eq:scaling_Ye_BHmass} (for a fiducial value of $\alpha_{\rm eff}$). Residual neutronization is due to the relaxation phase.

Based on the results discussed here, extrapolating the above scaling laws, we expect that the disk temperature is the main factor that determines the breakdown of the fiducial ignition threshold and thus of self-neutronization. We find based on Fig.~\ref{fig:Ye_trisim} that accretion flows around black holes of masses $\gtrsim 3000M_\odot$ are not able to self-neutronize anymore at the fiducial ignition threshold, leading to a termination of r-process nucleosynthesis at these black hole masses and accretion regimes. It is theoretically plausible that neutronization of such accretion flows may still proceed at $\dot{M}\gg \dot{M}_{\rm ign}$ for $M_\bullet \gtrsim 3000M_\odot$. Such scenarios need to be probed by additional simulations. Whether such extreme accretion regimes of $\dot{M}\gtrsim 100 M_\odot \,\text{s}^{-1}$ can be realized in nature also represents an open question.

\section{Conclusions}\label{sec:conclusions}

We have explored the conditions for super-collapsar accretion disks to neutronize in the black-hole mass range of $M_\odot \approx 80-3000 M_\odot$ by means of semi-analytic calculations as well as neutrino-cooled three-dimensional GRMHD simulations. The viscous-timescale GRMHD simulations of isolated collapsar accretion disks represent a controlled numerical setup that allows us to investigate the intrinsic physical conditions of accretion disks at and near the ignition threshold, independent of global and necessarily specific collapsar models. In contrast to earlier parametrized disk evolution studies \citep{Siegel_collapsars_2019,de_igniting_2021} that explore the accretion state at a certain value of $\dot{M}$, here we follow the accretion process over more than a viscous timescale to record in detail the physical conditions as the disks gradually transition through the ignition threshold from $\dot{M}\gtrsim \dot{M}_{\rm ign}$ to $\dot{M}\lesssim \dot{M}_{\rm ign}$. The semi-analytic collapsar model (Sec.~\ref{sec:Motivation}) links these intrinsic physical accretion conditions to concrete stellar models, i.e. the global context. Our conclusions can be summarized as follows.

\begin{enumerate}
    \item We analytically explore the assumptions underlying the scaling of the ignition threshold $\dot{M}_{\rm ign}\propto \alpha^{5/3}M_\bullet^{4/3}$ and their possible breakdown as the black-hole mass $M_\bullet$ increases by orders of magnitude (Sec.~\ref{sec:theory_ign_threshold}). These include (i) the dominance of radiation and $e^\pm$-pressure, (ii) optically thin accretion flows, (iii) cooling dominated by electron and positron capture reactions, and (iv) relativistic $e^\pm$. We find that (i) improves with increasing black-hole mass and that (ii) also likely holds indefinitely due to (i). A gradual breakdown of (iii) starting as several thousand $M_\odot$ will transition into a very similar scaling $\dot{M}_{\rm ign}\propto \alpha^{9/5}M_\bullet^{6/5}$ (Sec.~\ref{subsec:results_neutrino_cooling}). We predict a likely breakdown of (iv) due to a temperature decrease $T\propto M_\bullet^{-1/6}$ (Eq.~\eqref{eq:temperature_scaling}) starting at a few thousand $M_\odot$. 
    %\textcolor{red}{where does this high mass scaling relation come from? I would point to Sec. 5.5.3 or move that equation to methods in order to avoid confusion.}

    \item Our GRMHD simulations evolve super-collapsar accretion disks over more than an initial viscous timescale and explore the physical conditions as they transition through the ignition threshold, from an efficiently cooled, optically and geometrically thin state to an inefficiently cooled, advective, geometrically thick regime (Figs.~\ref{fig:rho_temperature_dual_var_image}, \ref{fig:nu_cooling_rates}). The viscously spreading disks approximately approach the expected $\dot{M}\propto t^{-4/3}$, $t_{\rm visc}\propto t$ behavior for advective disks (\citealt{metzger_time-dependent_2008}; Fig.~\ref{fig:accretion_tvisc}), showing that angular momentum transport via MHD turbulence and a quasi-stationary MHD dynamo is well captured (Figs.~\ref{fig:lambda_MRI_resolution} and \ref{fig:butterfly}). We present a general construction of a tetrad frame comoving with the accretion flow in an axisymmetric, asymptotocally flat spacetime (Appendix \ref{app:alpha_viscosity}) and extract a typical effective MHD $\alpha$-viscosity of $\alpha_{\rm eff}\approx 0.01$ near the ISCO in our accretion disks at the ignition threshold (Fig.~\ref{fig:alpha_visc}).

    \item Within diagnostic uncertainties the analytically expected scaling $\dot{M}_{\rm ign}\propto M_\bullet^{4/3}$ as well as $\dot{M}_{\rm ign}\propto \alpha^{5/3}$ is in remarkable agreement with our GRMHD simulation results over the entire range $M_\bullet \lesssim 3-3000 M_\odot$, i.e. from normal collapsar to super-collapsar accretion disks (Fig.~\ref{fig:Mdot_ign_comparison}). We track $\dot{M}_{\rm ign}$ in the simulations as the onset of electron degeneracy ($\langle \eta_e \rangle_{\hat D} = 0.5$) in the innermost accretion disk. The $M_\bullet = 3000 M_\odot$ model solely provides a lower limit on $\dot{M}_{\rm ign}$ and indicates the breakdown of ignition at the expected value of the scaling law owing to strong suppression of neutrino emission.
    %\textcolor{red}{$M_\bullet = 3000 M_\odot$ model approaches this threshold, but does not sufficiently exceed it (Fig. \ref{fig:eta_Ye}). (if not precisely this statement, I would add a sentence or two describing what we mean by the M = 3000 msun model not achieving ignition. See also the comment below.)} . 
    Equation \eqref{eq:ignition_threshold_formula_normalized_80Msun} presents our simulation calibrated scaling formula. 

    \item Associated with the onset of electron degeneracy ($\langle \eta_e \rangle \gtrsim 0.5$) is self-neutronization of the simulated GRMHD accretion disks (Fig.~\ref{fig:eta_Ye}). The average electron fraction in the inner part of the accretion disk at the ignition threshold increases as a function of black-hole mass, roughly as $\langle Y_e \rangle \propto M_\bullet^{1/6}$ (Fig.~\ref{fig:Ye_trisim}), a trend that we also analytically derive. 
    
    \item Given the optically thin conditions at $\dot{M}_{\rm ign}$ (Fig.~\ref{fig:electronneutrino_number_optical_depth}), $Y_e$ of the accreting plasma is also largely characteristic of the composition of the disk outflows. We find that for accreting black holes of $M_\bullet \lesssim 100 M_\odot$ at $\dot{M}\gtrsim \dot{M}_{\rm ign}$ the composition is sufficiently neutron-rich to synthesize significant amounts of lanthanides ($\langle Y_e \rangle_{\hat D} < 0.2$; Figs.~\ref{fig:rho_temperature_dual_var_image}, \ref{fig:eta_Ye}). We thus verify by explicit GRMHD simulations the assumptions of self-neutronization at and above the ignition threshold underlying the semi-analytic global collapsar-disk model of \citet{Siegel_super-kilonovae_2022} in the context of black holes populating the PISN mass gap. Bright and red super-kilonova transients that last months or longer as predicted by \citet{Siegel_super-kilonovae_2022} are indeed supported by the current GRMHD disk simulations. 

    \item The GRMHD simulations demonstrate the existence of a breakdown of the fiducial ignition threshold and associated self-neutronization at around $M_\bullet \gtrsim 3000 M_\odot$. Whereas the assumptions (i) and (ii) under 1.~are satisfied and improve in the case of (i) with increasing $M_\bullet$ (Figs.~\ref{fig:pressure_ratios}, \ref{fig:electronneutrino_number_optical_depth}), the transition into a similar scaling regime associated with the breakdown of (iii) is clearly above the mass range considered here (Fig.~\ref{fig:nu_cooling}). We associate the breakdown of the standard ignition scaling with the temperature decreasing as $T\propto M_\bullet^{-1/6}$ to $\lesssim\!1$\,MeV at $\gtrsim 3000 M_\odot$ (Fig.~\ref{fig:Ye_trisim}), which marks the neutron-proton mass difference. Whereas neutronization may still occur at $M_\bullet \gtrsim 3000M_\odot$, given the extreme accretion rates $\dot{M}\gtrsim 100 M_\odot \,\text{s}^{-1}$ needed to be realized in nature, this may remain merely a theoretical possibility. Our results suggest that for practical purposes self-neutronization and the synthesis of r-process elements may cease at black-holes of mass $\gtrsim 3000 M_\odot$. This limit is gradual in nature, however, because the neutrino cooling rates are skewed toward the high-energy tail of the Fermi-Dirac distribution for electrons and positrons, preventing a sharp cut-off of energetically significant cooling as a function of black-hole mass (Sec.~\ref{subsec:theo_validity_neutrino_cooling_rates}).  
    %\textcolor{red}{Now that you've added the n-p mass difference calculation which finds the threshold to be at M = 3900 msun, some readers might wonder at us saying $>$ 3000 so consistently. It might be worth adding a sentence somewhere couching this and reminding the reader how approximate this threshold is. You did a good job of this in the method section, but I am definitely guilty of sometimes reading the conclusion before the methods and others probably are as well.}

    \item Based on models of massive stars with low to vanishing metallicity and $M_{\rm ZAMS}\gtrsim 250-10^5M_\odot$ using the semi-analytic collapsar model of \citet{Siegel_super-kilonovae_2022} and our simulation-calibrated scaling formula for $\dot{M}_{\rm ign}$ (Eq.~\eqref{eq:ignition_threshold_formula_normalized_80Msun}), we find that black holes of $M_\bullet \sim 30-1000 M_\odot$ accreting at $\dot M \gtrsim \dot{M}_{\rm ign}$ can indeed be realized in nature (Fig.~\ref{fig:massive_Star_fallback}). We find that such super-collapsars may synthesize $\sim\!10-100 M_\odot$ of r-process material in disk outflows. If these ejecta escape the collapsar site, such super-collapsars may be very rare but highly prolific sources of both light and heavy r-process elements, specifically in low-metallicity environments.

    \item An intrinsic prediction of the scaling of physical conditions at the ignition threshold with black-hole mass (cf., e.g.,~Fig.~\ref{fig:Ye_trisim}) is the transition from significantly lanthanide bearing (`red') disk outflows for black holes in the range $M_\bullet \lesssim 100 M_\odot$ accreting at $\dot M \gtrsim \dot{M}_{\rm ign}$ to increasingly lanthanide poor (`blue') outflows and associated super-kilonova transients for black holes of $\gtrsim 500-1000 M_\odot$ accreting in their respective regime $\dot M \gtrsim \dot{M}_{\rm ign}$. A detailed nucleosynthesis analysis of outflows as well as predictions for lightcurves and spectra of the associated transients are beyond the scope of this paper (see \citealt{Siegel_super-kilonovae_2022}, however, for first models geared toward the lower mass end probed here). 
 
\end{enumerate}

The authors thank J.~Hernández Morales, M.~Müller, L.~Combi, and D.~Desai for discussions and comments. The authors gratefully acknowledge the computing time made available to them on the high-performance computer ``Lise'' at the NHR Center NHR@ZIB. This center is jointly supported by the German Federal Ministry of Education and Research and the state governments participating in the NHR (www.nhr-verein.de/unsere-partner). Support for the publication fee was provided by the University of Greifswald's publication fund. D.M.S.~acknowledges the support of the Natural Sciences and Engineering Research Council of Canada (NSERC), funding reference number RGPIN-2019-04684. B.D.M.~acknowledges support from the National Science Foundation (grant numbers AST-2002577, AST-2406637), and Simons Investigator Grant 727700. The Center for Computational Astrophysics at the Flatiron Institute is supported by the Simons Foundation.

\software{The Einstein Toolkit (\citealt{roland_haas_2020_4298887}; \href{http://einsteintoolkit.org}{http://einsteintoolkit.org}),  \texttt{GRMHD\_con2prim} (\citealt{siegel_grmhd_con2prim_2018}, \citealt{siegel_recovery_2018}), \texttt{WhiskyTHC} (\citealt{radice_thc_2012}, \citealt{radice2014HighOrder}, \citealt{radice2014Secondorder}),  \texttt{PyCactus} (\citealt{kastaun_numerical_2021}, \url{https://github.com/wokast/PyCactus}), \texttt{Matplotlib} \citep{hunter2007Matplotlib}, \texttt{NumPy} \citep{harris2020Array}, \texttt{SciPy} \citep{virtanen2020SciPy}, \texttt{hdf5} \citep{The_HDF_Group_Hierarchical_Data_Format}, and \texttt{h5py} \citep{Python_h5py}}

\appendix

\section{Construction of the comoving accretion flow tetrad}
\label{app:alpha_viscosity}

The magnetic and fluid stresses relative to the mean flow generate effective viscosity in the disk and lead to angular momentum transport and accretion onto the black hole. Here, we construct the tetrad vectors of the instantaneous comoving frame of the accreting fluid. The total stress tensor $t_{r\phi}$ is projected onto this frame to compute an effective $\alpha$-viscosity parameter $\alpha_{\rm eff}$ (Eq.~\eqref{eq:alpha_viscosity}). We build on the construction of tetrads for the comoving mean flow frame in Boyer-Lindquist coordinates (e.g.~\citealt{novikov_astrophysics_1973,2005_Krolik_Kerr_paper_alpha_tetrads,2008_Beckwith_Hawley_Krolik,2011Kulkarni_measuringBH_spin}) and generalize it to a general axisymmetric, asymptotically flat spacetime with metric $g_{\mu\nu}$ using spherical coordinates.

The tetrad vectors of a comoving observer $(e_{(t)},e_{(r)},e_{(\phi)},e_{(\theta)})$ are characterized by the property $e^\mu_{(a)}e^{(b)}_{\mu} = \delta^{(b)}_{(a)}$. The spacetime index $\mu$ is raised and lowered with the spacetime metric $g_{\mu\nu}$, whereas the tetrad component index $(a)$ is raised and lowered with the Minkowski metric $\eta_{(a)(b)} = \mathrm{diag}(-1,1,1,1)$. The transformations of a vector $X^\mu$ and co-vector $X_\mu$ into the comoving system are given by $X^{(a)} = e^{(a)}_\mu X^{\mu}$ and $X_{(a)} = e_{(a)}^\mu X_{\mu}$, respectively. The corresponding inverse transformations read $X^\mu = e^\mu_{(a)}X^{(a)}$ and $X_\mu = e_\mu^{(a)}X_{(a)}$.

Since the fluid flow is at rest in the co-moving frame, the only non-vanishing component of $u^\mu$ is the time component, and one has $e^{\mu}_{(t)} = u^\mu$ ($u^\mu$ is already adequately normalized). The remaining tetrad vectors are not uniquely defined. To recover the conventional meaning of the coordinate directions in the Newtonian and flat spacetime limit, we adopt the same ansatz as in \citet{2005_Krolik_Kerr_paper_alpha_tetrads}. Under the assumption of dominant azimuthal motion in the coordinate frame, $|u^\phi|\gg |u^r|,|u^\theta|$, we intend to align the comoving frame with the coordinate frame along the direction of largest difference and set the $r$ and $\theta$ components of $e^\mu_{(\phi)}$ to zero. The resulting expression \eqref{eq:e_phi_tetrad} is, modulo some misprints in \citet{2005_Krolik_Kerr_paper_alpha_tetrads}, identical to their expression, even for a general form of the metric. 

Subsequently, we derive the radial tetrad vector by setting the $\theta$-component to zero, i.e.~$e^{\mu}_{(r)} = (c,d,0,e)$, where $c$, $d$, and $e$ are three unknowns. The latter are computed by solving three coupled linear equations: two orthogonality relations regarding the $\phi$- and $t$-tetrads (i.e. $e^{\mu}_{(r)} e_{\mu}^{(\phi)} = g_{\mu\nu}e^{\mu}_{(r)}e^{\mu}_{(\phi)} = 0$ and $e^{\mu}_{(r)} e_{\mu}^{(t)} = -g_{\mu\nu}e^{\mu}_{(r)}e^{\mu}_{(t)} = 0$) and the normalization condition $e^{\mu}_{(r)} e_{\mu}^{(r)} = 1$. Finally, we construct the $\theta$-tetrad vector starting from a guess vector $\tilde{v}= (1,1,1,1)$, using the Gram-Schmidt procedure to obtain a vector orthonormal to the previously defined tetrad vectors. The final result for the tetrad vectors can be written as
\begin{eqnarray}
     e^{\mu}_{(t)} &=& u^\mu ,\\
     e^{\mu}_{(\phi)} &=& (g_{tt}l^2-2 g_{t\phi}  l+g_{\phi \phi} )^{-1/2} (-l,0,0,1), \label{eq:e_phi_tetrad}\\
     e^{\mu}_{(r)} &=& (c,d,0,e) ,\\
     e^{\mu}_{(\theta)} &=& \frac{v^\mu}{\sqrt{v^\nu v_\nu}},
\end{eqnarray}
%\begin{eqnarray}
where $l = \frac{u_{\phi}}{u_t}$ is the specific angular momentum and
\begin{equation}
    v^\mu \equiv \tilde{v}^\mu - (\tilde{v}_\nu e^{\nu}_{(t)})e^{\mu}_{(t)} - (\tilde{v}_\nu e^{\nu}_{(\phi)})e^{\mu}_{(\phi)} -(\tilde{v}_{\nu}e^{\nu}_{(r)})e^{\mu}_{(r)}.
\end{equation}
Setting
   \begin{eqnarray}
       \gamma_{1} =& g_{\phi \phi }u^2_{r}+g_{rr} u^2{}_{\phi },\\
        \gamma _2 =& g_{rr} g_{tt }-g^2_{tr}, \\
        \gamma _3 =& \frac{g_{tt } u_{\phi}^2}{u_{t }^2}+g_{\phi \phi },\\
        \gamma _4 =& g_{r \phi }-\frac{g_{t r } u_{\phi}}{u_{t }},\\
        \gamma _5 =& g_{t r } u_{r}-g_{rr} u_{t } ,\\
        \gamma _6 =& 2 l g_{t r }+g_{r \phi },\\ 
        \gamma _7 =& g_{tt } u_{r}-g_{t r } u_{t }, \\
       \gamma _8 =& g_{rr} g_{\phi \phi }-g^2_{r \phi }, \\
       \gamma_9 =& g_{rr} u_{\phi}-4 g_{r \phi } u_{r},
   \end{eqnarray}
and
   \begin{eqnarray}
       \xi_1 &=& g^2_{r \phi } g_{tt } u_{\phi}+\gamma _3 \gamma _7 g_{r \phi }-\gamma _2 g_{\phi \phi }u_{\phi},\\
       \xi_2 &=& g_{t r } lu_{\phi} u_{r}+u_{\phi} \left(-\gamma _9-g_{r \phi } u_{r}\right),\\
       \xi_3 &=& \gamma _9 u_{t }-\frac{g_{tt } u^2_{r} u_{\phi}}{u_{t }},\\
       \xi_4 &=& u^2_{t} \left(\gamma _8+l^2 g^2_{tr}\right)+\gamma _3 g_{tt } u^2_{r}-2 \gamma _3 g_{t r } u_{r} u_{t },\\
       \xi_5 &=& u_{t } \left(u_{\phi} \left(-\gamma _8-l g_{r \phi } g_{t r }\right)+g_{\phi \phi }u_{r} \gamma_6\right) +\gamma _4 \left(-g_{t r }\right) u^2{}_{\phi }+\gamma _5 g_{\phi \phi }u_{\phi},\\
       \xi_6 &=& g_{tt } u^2{}_{\phi } \left(g^2_{r \phi }+\gamma _2 l^2\right)+2 l g_{t\phi }^3 u^2_{r},\\
       \xi_7 &=& \gamma _{1}+2 lu_{\phi} \left(-\gamma _5-g_{t r } u_{r}\right),\\
       \xi_8 &=& g_{tt } lu_{\phi}+g_{\phi \phi }u_{t }-2 g_{t\phi } u_{\phi}, \\
       \xi_9 &=& g_{\phi \phi }u^2_{r}+g_{rr} u_{\phi}^2,\\
       \xi_{10} &=& l g_{t\phi } u_{r}-g_{\phi \phi }u_{r}+\gamma _4 u_{\phi},\\
       \xi_{11} &=& -l g_{tt } u_{r}+g_{t\phi } u_{r}-g_{r \phi } u_{t }+g_{t r } u_{\phi},
   \end{eqnarray}
we obtain the components of the $\theta$-tetrad vector as follows:
    \begin{eqnarray}
        c &=& \pm [\text{sgn}(\xi _8 )]^{-1}\xi _{10} \left\{g^2_{t \phi } \left(l \xi _3+\xi _7\right)+2 g_{t\phi } \left[l \left(\xi _2-\xi _9\right) g_{tt }+\xi _5\right]+\xi _4 g_{\phi \phi }+\xi _6-2 \xi _1 u_{\phi}\right\}^{-1/2},\label{eq:theta_tetrad_c}\\
        d&=& \left| \xi _8\right|  \left\{g^2_{t \phi } \left(l \xi _3+\xi _7\right)+2 g_{t\phi } \left[l \left(\xi _2-\xi _9\right) g_{tt }+\xi _5\right]+\xi _4 g_{\phi \phi }+\xi _6-2 \xi _1 u_{\phi}\right]^{-1/2},\label{eq:theta_tetrad_d}\\
        e &=& \left[\text{sgn}\left(\xi _8\right)\right]^{-1}\xi _{11} \left\{g^2_{t \phi } \left(l \xi _3+\xi _7\right)+2 g_{t\phi } \left[l \left(\xi _2-\xi _9\right) g_{tt }+\xi _5\right]+\xi _4 g_{\phi \phi }+\xi _6-2 \xi _1 u_{\phi}\right\}^{-1/2}\label{eq:theta_tetrad_e}.
    \end{eqnarray}
%\end{eqnarray}
 In the above equations, $\text{sgn}(x) = x/|x|$. In Eq.~\eqref{eq:theta_tetrad_c}, we choose the positive sign, such that $e^{r}_{(r)} > 0$, which ensures that the comoving tetrad has right-handed orientation. The above equations \eqref{eq:theta_tetrad_c}--\eqref{eq:theta_tetrad_e} are only valid solutions, provided the following condition is satisfied:
 \begin{eqnarray}
     g_{rr} &>& (\xi _8^2)^{-1}\left(g_{r \phi } \left(4 g^2_{t \phi } u_{r} u_{\phi}+g_{t\phi } \left(2 l^2 g_{t r } u^2_{t}-2 g_{\phi \phi }u_{r} u_{t }\right) +g_{tt } \left(2 u_{r} \left(\gamma _3 u_{\phi}-g_{t\phi } \left(2 l u_{\phi}+lu_{\phi}\right)\right)+\gamma _4 \left(2 l u_{t } u_{\phi}-u^2{}_{\phi }\right)\right)\right)\nonumber\right.\\
     &&+ u^2_{r} \left(g_{\phi \phi }\left(g^2_{t \phi }-l^2 g^2_{tt}\right)+g_{tt } \left(l^2 g^2_{t \phi }-g^2_{\phi \phi }\right)+2 l g_{tt } g_{t\phi } g_{\phi \phi }-2 l g_{t\phi }^3\right)+u^2_{t} \left(g_{\phi \phi }\left(g^2_{r \phi }-l^2 g^2_{tr}\right)-2 l g^2_{r \phi } g_{t\phi }\right)\nonumber\\
     &&+\left. g_{t r } \left(2 u_{r} \left(g^2_{\phi \phi } u_{t }+g_{t\phi } u_{\phi} \left(-\gamma _3+2 l g_{t\phi }-2 g_{\phi \phi }\right)+g_{tt } g_{\phi \phi }lu_{\phi}\right)-\gamma _4 \left(u^2{}_{\phi } \left(l g_{tt }-2 g_{t\phi }\right)+2 g_{\phi \phi }u_{t } u_{\phi}\right)\right)\right).
 \end{eqnarray}
We explicitly checked that this condition is satisfied in our domain, except for regions inside the ISCO, which are irrelevant for the computation of an effective $\alpha$-viscosity.

\begin{comment}
    
We calculate the angular momentum flux as:
 \begin{equation}
     j = \frac{T^{r}_{\phi}}{\dot{\rho}}
 \end{equation}
 where $\dot{\rho}$ is the density flux calculated from the continuity equation(as given in the valencia formulation of general relativistic magneto-hydrodynamics) as:
 \begin{equation}
     \dot{\rho} = \frac{\partial_r \left(\frac{\sqrt{\gamma} \alpha_l D u^r}{\omega_l}\right)}{\sqrt{\gamma}}
 \end{equation}

where $\gamma$ is the determinant of the spatial metric, $\alpha_l$ is the lapse, $w_l$ is the Lorentz factor,$u^r$ is the radial coordinate velocity,$u^r/\omega_l$ is the advection speed concerning radial coordinate, and $D=\sqrt{\gamma}\rho\omega_l$ is the conserved density. We only account for $\dot{\rho}$ only in the radial direction and restrict the divergence to $\partial_r$ and ignore the angular directions.
\end{comment}
%\include{appendices/Derivation_of_weak_interaction_scaling}
%\include{appendices/Benchmarking_magnetic_fields}

%% For this sample we use BibTeX plus aasjournals.bst to generate the
%% the bibliography. The sample631.bib file was populated from ADS. To
%% get the citations to show in the compiled file do the following:
%%
%% pdflatex sample631.tex
%% bibtext sample631
%% pdflatex sample631.tex
%% pdflatex sample631.tex

\bibliography{references,DanielBib,gr-astro}
\bibliographystyle{aasjournal}

%% This command is needed to show the entire author+affiliation list when
%% the collaboration and author truncation commands are used.  It has to
%% go at the end of the manuscript.
%\allauthors

%% Include this line if you are using the \added, \replaced, \deleted
%% commands to see a summary list of all changes at the end of the article.
%\listofchanges

\end{document}